\documentclass[12pt,preprint]{aastex}

\def\simgt{\lower 2pt \hbox{$\, \buildrel {\scriptstyle >}\over{\scriptstyle \sim}\,$}}

\def\simlt{\lower 2pt \hbox{$\, \buildrel {\scriptstyle <}\over{\scriptstyle \sim}\,$}}

\begin{document}
\title{X-ray Absorption and an X-ray Jet in the Radio-Loud Broad
Absorption Line Quasar PG 1004+130}

\author{B.P. Miller,\footnotemark[1] ~W.N. Brandt,\footnotemark[1]
~S.C. Gallagher,\footnotemark[2] ~A. Laor,\footnotemark[3]
~B.J. Wills,\footnotemark[4] 
~G.P. Garmire,\footnotemark[1] ~\& D.P. Schneider\footnotemark[1]}

\footnotetext[1]{Department of Astronomy and Astrophysics, The
Pennsylvania State University, 525 Davey Laboratory, University Park,
PA 16802, USA; {\it bmiller, niel, garmire, dps@astro.psu.edu}}

\footnotetext[2]{Department of Physics and Astronomy, University of
California - Los Angeles, 430 Portola Plaza, Box 951547, Los Angeles
CA, 90095-1547, USA; {\it sgall@astro.ucla.edu}}

\footnotetext[3]{Department of Physics, Technion-Israel Institute of
Technology, Haifa, 32000, Israel; {\it laor@physics.technion.ac.il}}

\footnotetext[4]{Astronomy Department, University of Texas - Austin, 1
Univ. Station, C1400, Austin, TX 78712-0259; {\it
bev@astro.as.utexas.edu}}

	
\begin{abstract}

We investigate the X-ray properties of PG~1004+130, a radio-loud broad
absorption line (BAL) quasar with a hybrid FR~I/FR~II radio
morphology. This optically bright, low-redshift quasar was undetected
by {\it Einstein}, marking it as anomalously X-ray weak relative to
other radio-loud quasars. The 22.2~ks {\it XMM-Newton} and 41.6 ks
{\it Chandra} observations presented here are the first \hbox{X-ray}
detections of PG~1004+130 and constitute the highest spectral quality
X-ray observations of a radio-loud BAL quasar available to date. The
{\it Chandra} \hbox{ACIS-S} spectrum shows evidence for complex soft
X-ray absorption not detected in the data obtained 1.7 yr previously
with {\it XMM-Newton}, with a best-fit intrinsic column density of
$N_{\rm H}$=1.2~$\times$~10$^{22}$~cm$^{-2}$ for the preferred
partial-covering model. There is no significant difference in the
hard-band power-law photon index of $\Gamma\approx$1.5 between the two
observations. The {\it Chandra} image also reveals extended X-ray
emission $\approx$8$''$ (30 kpc) south-east of the nucleus, aligned
with the FR~I jet but upstream of the 1.4 GHz radio-brightness
peak. The jet is not detected by {\it HST}, and the optical upper
limit rules out a simple single-component synchrotron interpretation
of the radio-to-X-ray emission. The multiwavelength characteristics of
the PG~1004+130 jet, including its relatively flat X-ray power law and
concave spectral energy distribution, are similar to those of powerful
FR~II \hbox{jets}. The lack of strong beaming in PG~1004+130 limits the
efficiency of inverse Compton upscattering, and we consider the X-ray
emission to most likely arise from a second synchrotron component
generated by highly energetic electrons.

\end{abstract}


\section{Introduction}

Broad absorption line (BAL) quasars show deep and wide blueshifted
absorption troughs in their rest-frame UV spectra (e.g., Weymann et
al.~1991). The intrinsic fraction of quasars with BALs is $\approx$
20\% (e.g., Hewett \& Foltz~2003); in the most commonly accepted
scenario this represents the covering factor of an outflowing BAL
wind. While all BAL quasars show absorption from high-ionization
transitions, such as Si IV and C IV, some also display absorption from
lower ionization transitions, such as Mg II, and such objects tend to
be more reddened (e.g., Sprayberry \& Foltz~1992; Reichard et
al.~2003).

Although BALs were once believed to be confined to radio-quiet
sources, optical spectroscopy of the quasars from the {\it VLA} 1.4
GHz FIRST survey \citep {1995ApJ...450..559B}, conducted by the FIRST
Bright Quasar Survey (White et al.~2000), and the Sloan Digital Sky
Survey (SDSS; York et al.~2000) has revealed many radio-loud BAL
quasars (e.g., Becker et al.~2000; Menou et al.~2001). Radio-loudness
is commonly parameterized by the ratio of rest-frame radio-to-optical
flux densities, $R^{*}~=~f_{\rm 5 GHz}/f_{\rm 2500 A}$
\citep{1980ApJ...238..435S}, where those quasars with
$\log{R^{*}}~>$~1 are considered radio-loud. As this definition may
somewhat exaggerate the radio-loudness of BAL quasars with substantial
optical reddening, it is also useful to consider the rest-frame radio
luminosity density, where $L_{\rm 5 GHz}~>$~10$^{32}$ erg~s$^{-1}
$~Hz$^{-1}$ distinguishes radio-loud quasars (e.g., Miller et
al.~1993). The majority of radio-loud BAL quasars discovered to date
meet both criteria but tend to be of intermediate radio-loudness
(fewer than 20\% in the combined samples of Becker et al.~2000 and
Menou et al.~2001 have $\log{R^{*}}~>$~2). \citet{2000ApJ...538...72B}
found that about one-third of their BAL quasars were flat-spectrum
radio sources (some radio loud), suggesting a viewing angle for this
subset well above the equatorial region where disk-associated BAL
outflows would be visible, and further discovered that most of their
BAL quasars were compact (80\% unresolved to 0.2$''$ at 8~GHz). These
results support alternate proposals that BALs may be associated with
an evolutionary phase with a large BAL wind covering fraction, rather
than orientation (e.g., Gregg et al.~2000), although the consistency
of the sub-millimeter emission of BAL quasars with that of non-BAL
quasars is difficult to reconcile with evolutionary scenarios in which
BAL quasars are emerging from a dusty ``shroud'' (Lewis et al.~2003;
Willott et al.~2003).

Radio-quiet BAL quasars, particularly those with low-ionization
absorption features, are usually weaker in X-rays than would be
expected from their optical luminosities (e.g., Green et al.~2001;
Gallagher et al.~2006).  The X-ray spectra of radio-quiet BAL quasars
show clear evidence of X-ray absorption, often complex, with intrinsic
column densities \hbox{$N_{\rm H} >$ 10$^{22}$ cm$^{-2}$} (e.g.,
Gallagher et al.~2002).  Although UV and X-ray absorption are clearly
linked (e.g., Brandt, Laor, \& Wills 2000), the higher column density
of the X-ray absorber (e.g., Gallagher et al.~2006) suggests the X-ray
absorption arises interior to the UV BALs, perhaps in the ``shielding
gas'' postulated by Murray et al.~(1995) and generated naturally in
the simulations of \citet{2000ApJ...543..686P}. A {\it Chandra}
snapshot survey of five radio-loud BAL quasars confirmed they are also
X-ray weak relative to similar non-BAL quasars, but with fairly soft
spectra incompatible with the large column densities necessary to
explain this weakness as simple neutral absorption (Brotherton et
al.~2005).

Many radio-loud quasars display striking extended emission in the form
of jets and lobes. The classification scheme of Fanaroff \& Riley
(1974) distinguishes the edge-darkened, core-dominated FR~I sources
from the more luminous edge-brightened, lobe-dominated FR~II
objects. The multiwavelength properties of FR~I and FR~II jets are
quite distinct: FR~I jets have linear or convex spectral energy
distributions (SEDs) that can typically be modeled by a synchrotron
spectrum extrapolated (with a break if necessary) from the radio
through the optical to the X-ray (e.g., Worrall et al.~2004), while
the SEDs of the prominent knots in FR~II jets are concave, frequently
possessing only an upper limit at optical wavelengths (e.g., Sambruna
et al.~2004; Marshall et al.~2005). The X-ray emission in FR~II jets
is often interpreted as arising from inverse-Compton scattering of
cosmic microwave background photons (IC/CMB models; e.g., Tavecchio et
al.~2000; Celotti et al.~2001), a process that is increasingly
efficient for highly relativistic bulk motions, for jet angles close
to the line of sight, and for objects at large redshifts (Schwartz
2002). Some observational and theoretical complications with IC/CMB
models have been noted (e.g., Tavecchio et al.~2003; Hardcastle 2006),
and high-redshift radio-loud quasars in general do not appear to have
extreme jet-to-core X-ray brightness ratios (e.g., Siemiginowska et
al.~2003; Bassett et al.~2004; Lopez et al.~2006). Alternative models
for the origin of the X-ray emission from FR~II jets have been
proposed, including synchrotron emission from a population of highly
relativistic electrons (e.g., Atoyan \& Dermer 2004). X-ray hotspots
have been detected by {\it Chandra} in a number of FR~II sources; the
most luminous hotspots are consistent with synchrotron self-Compton
(SSC) emission but other cases require a second highly energetic
synchrotron component to produce the X-ray emission (Hardcastle et
al.~2004).

PG~1004+130 (PKS 1004+13, 4C 13.41) is an optically bright
($V$~=~14.98; e.g., Garcia et al.~1999), radio-loud
($\log{R^{*}}~=$~2.32 [Wills et al.~1999] and $L_{\rm 5
GHz}~=$~6.5$\times$10$^{32}$ erg~s$^{-1} $~Hz$^{-1}$), lobe-dominated
quasar at a redshift of $z$~=~0.240
\citep{2004ApJS..150..181E}. PG~1004+130 is notably X-ray weak; it was
undetected by {\it Einstein} with a 0.5--4.5~keV flux limit of 1.4
$\times$ 10$^{-13}$ erg cm$^{-2}$ s$^{-1}$
\citep{1984ApJ...280...91E}. \citet{1999ApJ...520L..91W} analyzed {\it
IUE} and {\it HST} (Goddard High Resolution Spectrograph; GHRS)
spectra of PG~1004+130 and found evidence for broad, blueshifted
absorption in several UV lines, notably O~{\rm VI}, N~{\rm V}, and
C~{\rm IV}. They concluded that PG~1004+130 was likely a BAL quasar
with a BALnicity index\footnotemark[1] (Weymann et al.~1991) of
$\approx$ 850 km s$^{-1}$. \footnotetext[1]{BALnicity index is
calculated from the C IV BAL as the total equivalent width of all
associated absorption troughs blueshifted from the emission peak by
greater than 3000 km s$^{-1}$ and at least 2000 km s$^{-1}$ wide. Any
quasar with a BALnicity index $>$~0~km~s$^{-1}$ is formally a BAL
quasar according to the criterion of Weymann et al.~(1991).}A recent
observation with {\it HST} performed with the Space Telescope Imaging
Spectrograph (STIS) to search for BAL variability has confirmed this
designation (B.~J.~Wills, in preparation), making PG~1004+130 the only
currently known low-redshift ($z <$ 0.5) radio-loud BAL quasar. These
results suggest that the X-ray weakness of PG~1004+130 is likely
related to absorption.

The extended radio structure of PG~1004+130 is also notable: it is one
of the prototypical examples discussed by Gopal-Krishna \& Wiita
(2000) of a hybrid morphology source possessing both an FR~I and an
FR~II lobe. The majority of the radio emission in the FR~I south-east
(SE) structure is concentrated in a knotted jet that is most prominent
close to the nucleus and then fades into a gradually broadening plume,
whereas the north-west (NW) structure shows classical FR~II
edge-brightened morphology. The structure of such hybrid sources is
more intuitively explained as arising from the propagation of twin
jets into dissimilar large-scale environments rather than by invoking
a ``lopsided'' central engine (Gopal-Krishna \& Wiita 2000). Both the
extended projected size of PG~1004+130 ($\sim$500 kpc) and the ratio
of lobe-to-core flux (we measure this to be $\sim$40 at rest-frame 1.7
GHz using FIRST data) suggest the jet axis is inclined at a relatively
large angle to our line of sight, consistent with physical models in
which BAL winds flow equatorially from the accretion disk. A
quantitative lower limit on the inclination to the line-of-sight,
$\theta$, may be estimated from the core radio-to-optical luminosity
ratio, $\log{R_{V}}~=~\log{(L_{\rm core}/L_{\rm opt})}$, which Wills
\& Brotherton (1995) find improves upon the commonly used lobe-to-core
ratio. PG~1004+130 has $\log{R_{V}}~\sim$~1.1, suggesting
$\theta~\simgt$~45$^{\circ}$ (consistent with Wills et al.~1999). This
large inclination angle implies little line-of-sight beaming in the
jet and constrains the interpretation of any associated X-ray
emission.

We have obtained the first X-ray detections and spectra of PG~1004+130
in order to check for nuclear absorption and determine the quasar's
general X-ray properties. We make use of both {\it XMM-Newton} and
{\it Chandra}, as they provide complementary information: {\it
XMM-Newton} has high throughput and covers a broader energy range,
whereas {\it Chandra} provides the angular resolution necessary to
detect and characterize any extended X-ray emission. The two X-ray
observations were conducted $\sim$1.7 years apart, allowing
investigation of both short and long-term X-ray variability.

In this paper we adopt a standard cosmology with $H_{0}$ = 70 km
s$^{-1}$~Mpc$^{-1}$, ${\Omega}_{M}$~=~0.3, and
${\Omega}_{\Lambda}$~=~0.7 \citep{2003ApJS..148..175S}. This choice
results in a luminosity distance of 1200 Mpc and an angular distance
scale of 3.8 kpc arcsec$^{-1}$ for PG~1004+130. The Galactic column
density toward PG~1004+130 (${\alpha}_{2000}$~=~10~07~26.10,
${\delta}_{2000}$~=~+12~48~56.20) is 3.93$\times$10$^{20}$ cm$^{-2}$
\citep{1996ApJS..105..369M}. Unless otherwise noted, errors are given
as 90\% confidence intervals for one parameter of interest
(${\Delta}{\chi}^{2}$~=~2.71).

\section{Observations}

PG~1004+130 was observed by {\it XMM-Newton}
\citep{2001A&A...365L...1J} on 2003 May~4 (ObsID 0140550601) for an
effective exposure time of 22.2~ks, and data were collected by all
instruments. There were insufficient counts in the RGS grating spectra
for meaningful analysis; we therefore restrict our study of the X-ray
spectrum to the EPIC (pn and MOS) imaging spectroscopy data. There was
no flaring during the observation in the detector-wide count rate for
events with energies greater than 10~keV.

Analysis was performed with the {\it XMM-Newton} Science Analysis
Software (SAS~v6.5.0). The nuclear spectrum was extracted from a
circular region with a radius of 24.6$''$, while background regions
were determined separately for the pn (a 533$''$ by 82$''$ rectangle)
and MOS (a 381$''$ by 192$''$ rectangle) detectors to avoid detector
gaps. We generated custom redistribution matrix files (RMFs) and
ancillary response files (ARFs), which together describe the
instrument response to incident photons and allow forward-fitting of
the spectrum. Finally, the extracted nuclear spectrum was binned to
contain at least 20 counts per bin using the FTOOLS task GRPPHA.
There are $\sim$1550 counts from 0.5--8~keV in the pn spectrum, and
$\sim$1100 total counts in the MOS1 and MOS2 spectra; the expected
number of background counts is $\sim$50 in the pn spectrum and
$\sim$30 total in the MOS spectra.

The observation included sets of images from the {\it XMM-Newton}
Optical Monitor (OM) for each of the UVW1, UVM2, and UVW2 filters,
with central wavelengths of 2910~\AA, 2310~\AA, and 2120~\AA,
respectively. We converted background-subtracted OM count rates
(automatically calculated during pipeline processing) to flux
densities using scaling factors appropriate for active galactic nuclei
(AGNs), as determined by the OM Calibration Team. The resulting UVW1,
UVM2, and UVW2 observed fluxes for PG 1004+130 are 6.3, 8.2, and 8.1
$\times$ 10$^{-15}$ erg~cm$^{-2}$~s$^{-1}$~{\AA}$^{-1}$,
respectively. The standard deviation of the individual flux density
measurements in a given filter is $\simeq$ 0.2 $\times$ 10$^{-15}$
erg~cm$^{-2}$~s$^{-1}$~{\AA}$^{-1}$, while the systematic uncertainty
in the flux calibration is less than 10\%.\footnotemark[2]
\footnotetext[2]{OM Calibration Team:
http://xmm.vilspa.esa.es/sas/documentation/watchout/uvflux.shtml}

We obtained a subsequent {\it Chandra} observation to confirm the
X-ray spectral properties, to investigate variability, and to examine
(at higher angular resolution) any extended \hbox{X-ray}
emission. PG~1004+130 was observed by {\it Chandra} on 2005 Jan~5
(ObsID 5606) for 41.6~ks with the Advanced CCD Imaging Spectrometer
(ACIS; Garmire et al.~2003) S-array in faint mode. These data have
been reprocessed (REPRO-III) to incorporate the latest calibration
(CALDB v3.2.0) including automatic application of both the ACIS
charge-transfer inefficiency correction and the time-dependent gain
adjustment. The ACIS-S3 image shows a bright point source at the
location of the nucleus of PG~1004+130, jet-associated X-ray emission
collinear with but mostly upstream of the SE radio jet, and several
additional sources, two of which are coincident with the outer regions
of the SE FR~I lobe. These two sources have point-source {\it HST}
counterparts; utilizing the methodology and results of Maccacaro et
al.~(1988), we find that their X-ray-to-optical flux ratios are
consistent with those expected for AGNs. We extracted the nuclear and
the extended X-ray emission for spectral analysis, using regions as
shown in Figure~1a (a circle with a 4.8$''$ radius for the nucleus and
a 4.1$''$ by 6.3$''$ rectangle for the jet) and measuring the
background from a large source-free elliptical region (with a 91$''$
major axis and a 38$''$ minor axis) east of the nucleus. There was no
significant background flaring during the observation. Source-specific
RMFs and ARFs were generated from the CALDB v3.2.0 database; the ARFs
include the effects of low-energy quantum-efficiency degradation due
to contamination build-up on the ACIS optical-blocking filter. The
extracted nuclear spectrum was binned to contain at least 20 counts
per bin using the FTOOLS task GRPPHA. The 0.5--8~keV nuclear spectrum
contains $\sim$1800 counts (of which only $\approx$~5 are expected to
be from background), and there are $\sim$30 counts in the extended
X-ray emission.

We supplement the {\it XMM-Newton} and {\it Chandra} data with
previously published and archival radio, IR, optical, and UV coverage
of PG~1004+130. {\it VLA} observations of PG~1004+130 have been
performed in several configurations and bands, and we make use of 1.4
GHz, 5.4$''$ FWHM catalog images from the FIRST survey \citep
{1995ApJ...450..559B} as well as C band (B configuration) data from
1979 taken by E.~Fomalont (Fomalont 1982) and L band (A configuration)
data from 1982 taken by J.~Wardle. Images for these latter two
observations were constructed using $uv$ components between 20 and 160
$k\lambda$, allowing investigation of the small-scale structure along
the jet at a resolution of 1.5$''$ while excluding the large-scale
structure already apparent in the FIRST image. IR fluxes were obtained
from {\it IRAS} (100--12 $\mu$m), ground-based ($N$, $M$, and $L$
bands), and {\it 2MASS} ($K$, $H$, and $J$ bands) data. An optical
spectrum and photometric $ugriz$ magnitudes were taken from the SDSS
database (PG~1004+130 is SDSS J100726.10+124856.2) and an {\it HST}
WFPC2 image of PG~1004+130, taken with the F606W filter and analyzed
and discussed by Bahcall et al.~(1997), was used to search for
extended optical emission. PG~1004+130 has been observed
spectroscopically in the UV by {\it HST} (initially with GHRS and
later with STIS; Wills et al.~1999; B.~J.~Wills, in preparation), {\it
IUE} (with both the long and short-wavelength spectrographs), and {\it
FUSE}, and portions of these data have been incorporated as well.

X-ray, optical, and radio images of PG~1004+130 are presented in
Figure 1. The \hbox{0.5--4~keV} X-ray image of Figure 1a has been
constructed with ACIS pixel randomization removed to increase angular
resolution and is shown with overlaid 4.9~GHz {\it VLA} contours; the
jet-associated X-ray emission commences upstream of the radio jet and
extends to the first radio knot. Examination of the optical {\it HST}
image with accompaning radio and smoothed X-ray contours reveals no
apparent optical counterpart to the SE FR~I jet. A spectral index map
of PG~1004+130 generated from 4.9 and 1.5 GHz {\it VLA} images of
$\sim$1.5$''$ resolution is presented in Figure 1c, along with a
high-resolution (0.5$''$ FWHM) 5~GHz {\it VLA} image of the nuclear
region. There does not appear to be any extended kpc-scale inner-jet
radio emission from PG~1004+130. The spectral index ${\alpha}_{\rm r}$
is $-$0.41 for the nucleus ($f_{\nu}\propto{\nu}^{\alpha}$) and
varies along the jet. The error in the spectral index increases
greatly toward the edges of the emission, but the signal-to-noise
ratio in both bands is quite high down the spine of the jet; the
repeated spectral steepening is genuine. We rebinned the X-ray image
to 0.05$''$ pixels and extracted the radial profile along the SE radio
jet (using concentric annular sectors restricted to a 30$^{\circ}$
arc) as well as excluding the jet (using annular sectors covering the
remaining 330$^{\circ}$), subtracting background as determined from a
nearby rectangular region. The background-subtracted
surface-brightness profiles are compared with normalized ACIS-S point
spread functions (created with MKPSF) in Figure 1d. As the nucleus
contains $\sim$1080 counts from 0.5--2 keV and $\sim$730 counts from
2--8 keV, the pre-calculated 1.5 keV response data are sufficient for
this analysis. The nucleus appears unresolved, but the jet-associated
emission is inconsistent with a point-source profile.

We are confident that the jet-associated X-ray emission arises in a
genuine counterpart to the SE FR~I radio jet due to the low
probability of finding a background source so close to the PG~1004+130
nucleus, the high X-ray-to-optical flux ratio of this feature, and the
extended nature of the X-ray emission as well as its position upstream
of the peak radio brightness in the jet. From Bauer et al.~(2004), the
expected sky density of sources (primarily AGNs) with 0.5--2~keV X-ray
fluxes comparable to or greater than the \hbox{$S_{0.5-2} \approx$
1.8$\times$10$^{-15}$ erg~cm$^{-2}$~s$^{-1}$} of the jet is $N(>S)
\approx$~500~deg$^{-2}$. Consequently, the probability of finding a
background source of appropriate brightness within 10$''$ and aligned
(within 30$^{\circ}$) with either the jet or counterjet of PG~1004+130
is only $\approx$~0.002. Further, the lack of any {\it HST} point
sources coincident with the X-ray emission restricts the nature of any
such background source; the optical upper limit yields an
X-ray-to-optical flux ratio, as defined by Maccacaro et al.~(1988),
that is more than $\sim$20 ($\sim$8) times greater than the maximum
value for AGNs (BL Lacs) observed by Maccacaro et al.~(1988). The
X-ray emission is extended along the path of the radio jet, and the
peak X-ray brightness in the jet occurs upstream of the peak radio
brightness, consistent with the observed trend for the ratio of
X-ray-to-radio emission to decrease along jets (e.g., Marshall et
al.~2002; Sambruna et al.~2004).

Large-scale diffuse X-ray emission is common within groups and
clusters, is sometimes seen in halos around RLQs, and has been
observed along radio jets. The intra-group, intra-cluster, and halo
X-ray emission is well-established to be thermal radiation from hot
gas, but the origin of diffuse X-ray emission along jets is much less
certain; IC/CMB scattering has been suggested as a reasonable
explanation (e.g., Siemiginowska et al.~2002; Schwartz et
al.~2005). We adaptively smoothed the 0.5--4~keV {\it Chandra} image
to search for diffuse extended emission in the vicinity of
PG~1004+130; the resulting image is presented in Figure 2a with
overlaid 1.4~GHz FIRST contours. The local environment of PG~1004+130
is sparsely populated; McLure \& Dunlop (2001) find that PG~1004+130
has the lowest clustering amplitude among the 13 RLQs they surveyed,
and the nearest known companion galaxy lies 33.4$''$ (127 projected
kpc) to the south-west, with another associated galaxy 45$''$ (170
kpc) to the west (Stockton 1978; Bahcall et al.~1997). There does not
appear to be any detectable X-ray halo of quasi-spherical, hot,
thermally emitting gas around PG~1004+130. The hybrid radio morphology
of PG~1004+130 may suggest that the density of the surrounding medium
is higher to the SE, where the radio jet is quickly decollimated, than
to the NW, where the unseen counterjet powers a distant,
edge-brightened lobe, although the lack of any apparent interaction
with the few neighboring field galaxies leaves the mechanism for
generating such an asymmetrical density distribution unclear. We do
not observe any strong gradient in diffuse X-ray emission from the SE
to the NW, but naturally this does not prohibit large-scale,
asymetrically distributed, colder gas. However, PG~1004+130 does show
intriguing diffuse emission along the path of the SE jet and NW
counterjet. We verified that this emission was not an artifact of the
smoothing process by extracting radial profiles from the rebinned
image as with the PSF analysis described previously; here we use the
regions indicated in Figure 2a to obtain surface-brightness profiles
along the jet, counterjet, and non-jet background (excluding all point
sources). Figure 2b shows that beyond $\sim$16$''$ (past the SE X-ray
jet emission), the diffuse emission in the jet and counterjet is
similar and clearly surpasses the emission from the non-jet background
region; we calculate $\sim$29.3 counts above background in the 0.5--4
keV diffuse emission from 16$''$ to 63$''$ to the SE, and $\sim$27.4
to the NW. Note that the ACIS readout streak cannot contribute
significantly to the diffuse emission, as it should contain less than
6 total counts over the jet/counterjet region, is offset by
$\sim$15$^{\circ}$ from the jet position angle, and would extend the
entire length of the chip. Extended X-ray emission around RLQs has
occasionally been noted to follow the direction of the radio lobes
(e.g., Croston et al.~2004; Belsole et al.~2006) but detection of an
X-ray counterjet component is somewhat unusual (cf. Schwartz et
al.~2005). We briefly consider the nature of this large-scale emission
in $\S$4.2.

The X-ray light curves of PG~1004+130 do not reveal significant rapid
variability within either the {\it XMM-Newton} or the {\it Chandra}
observations. The cumulative photon arrival times are consistent with
a constant count rate with Kolmogorov-Smirnov probabilities of 0.64
and 0.61 for {\it XMM-Newton} (pn detector) and {\it Chandra},
respectively. The 1$\sigma$ upper limit to the \hbox{0.5--8}~keV
count-rate variability on 1~ks timescales is $<$12\% over the {\it
XMM-Newton} observation, and $<$15\% over the {\it Chandra}
observation. After accounting for the relative number of soft
(0.5--2~keV) and hard (2--8~keV) band counts, the 1$\sigma$ upper
limit to hard band variability is somewhat greater than that for the
soft band.

\section{X-ray Spectral Analysis}

We analyzed the X-ray spectra using the most recent (v12.2.1) XSPEC
(Arnaud 1996) software package. Data associated with energies below
0.4 keV or above 8.0 keV were discarded for the purposes of fitting;
these cutoffs were imposed due to increasing calibration uncertainties
at low energies and declining source counts (due to decreasing
instrumental effective area and the spectral shape) as well as
increasing background at high energies. All fits include Galactic
absorption fixed at 3.93$\times$10$^{20}$ cm$^{-2}$
\citep{1996ApJS..105..369M}. Joint fitting of the spectra from the
{\it XMM-Newton} pn, MOS1, and MOS2 detectors increases the
signal-to-noise and is consequently preferred in the absence of
significant cross-calibration uncertainties. Fitting a simple
power-law model gives consistent results (the 90\% confidence
intervals for both photon index and normalization overlap for the pn,
MOS1, and MOS2 detectors), therefore all model parameters with the
exception of normalization have been fit jointly throughout the
remainder of the analysis. Unless otherwise noted we quote the outcome
of fitting various models to the binned spectra with ${\chi}^{2}$
minimization, but the following results agree with those obtained by
fitting the ungrouped spectra with the XSPEC $C$-statistic (after Cash
1979).

Given the BALs observed in the UV by \citet{1999ApJ...520L..91W} and
the correlation between BALs and X-ray absorption (e.g., Gallagher et
al.~2002), we anticipated potential X-ray absorption at low
energies. In an effort to disentangle the underlying power-law from
any intrinsic absorption, we initially fit only the data in the 2--8
keV range with a power-law model. The photon index obtained by jointly
fitting the pn, MOS1, and MOS2 data from {\it XMM-Newton} is
$\Gamma$~=~1.57$_{-0.19}^{+0.13}$, consistent with the {\it Chandra}
result of $\Gamma$~=~1.52$_{-0.26}^{+0.16}$. This photon index is
typical for the high-energy spectra of radio-loud quasars (RLQs);
e.g., Reeves \& Turner (2000) found $\langle{\Gamma}\rangle$~=~1.66
with $\sigma$~=~0.22 for an {\it ASCA} sample of 35 RLQs, while Page
et al.~(2005) determined $\langle{\Gamma}\rangle$~=~1.55 with
$\sigma$~=~0.29 for an {\it XMM-Newton} sample of 16 RLQs at
$z>~$2. The 2--8~keV {\it XMM-Newton} pn model flux is
2.57$^{+0.35}_{-0.66}\times$10$^{-13}$ erg~cm$^{-2}$~s$^{-1}$, 23\%
lower than the {\it Chandra} result of
3.32$^{+0.87}_{-1.05}\times$10$^{-13}$ erg~cm$^{-2}$~s$^{-1}$; this
flux difference exceeds cross-calibration uncertainties, which are on
the order of 12\% (S. Snowden 2005, private communication).

Many AGNs with significant X-ray absorption also have strong Fe
K$\alpha$ emission lines (e.g., see the {\it ASCA} observations of
Seyfert 2 galaxies by Turner et al.~1997). This was predicted by
Krolik \& Kallman (1987), who argued that the prominence of iron
features produced in the scattering and reflection regions increases
as the observed X-ray continuum becomes dominated by reprocessed
radiation. Examination of the residuals from the power-law fit
described above did not show any obvious emission features in the
X-ray spectrum of PG~1004+130. We tested for Fe K$\alpha$ emission by
adding an unresolved ($\sigma$~=~0.01~keV) redshifted Gaussian
emission line to the 2--8~keV power-law model. Here we fit the
ungrouped data (with the XSPEC $C$-statistic), as the large energy
bins at high energy would tend to smooth out any narrow line in the
binned spectra. The {\it XMM-Newton} pn and MOS spectra show no
evidence for neutral or ionized Fe K$\alpha$ emission and an upper
limit of 105~eV (90\% confidence) can be placed on the rest-frame
equivalent width of a narrow line at a fixed rest-frame energy of
6.4~keV from the pn spectrum. Modeling the {\it Chandra} spectrum by
adding a narrow line at rest-frame 6.4~keV imposes an upper limit of
121~eV on the intrinsic equivalent width. If the line energy is
allowed to vary, there is marginal evidence for emission at
6.57$^{+0.15}_{-0.11}$ keV; the rest-frame equivalent width of the
line is only 105$^{+109}_{-88}$ eV. Since this line energy does not
match the transition energy for Fe K$\alpha$ emission from either
neutral/low ionization (6.4~keV), He-like (6.7~keV), or H-like
(6.9~keV) iron, if the marginal detection is assumed to be indicative
of physical conditions then there must be a range of ionization states
in the scattering material. Overall, the lack of strong Fe K$\alpha$
emission argues against a scenario in which the nucleus of PG~1004+130
is heavily absorbed and the weak continuum arises chiefly from
scattering or reflection.

We extrapolated the 2--8~keV power-law fits to lower energies to
search for X-ray absorption, as illustrated in Figures 3a and 3b. Many
quasars, including radio-loud objects, show enhanced flux above a
power-law model below $\sim$2~keV (Porquet et al.~2004; Brocksopp et
al.~2006); no such soft excess is observed from PG~1004+130. While
there is only minimal evidence for intrinsic absorption in the {\it
XMM-Newton} spectra, the large systematic negative residuals in the
{\it Chandra} spectrum indicate substantial absorption. This
significant change in the absorption properties of PG~1004+130
occurred over only 494 rest-frame days. We characterized this
absorption by adding a redshifted neutral absorber to our model and
expanding the range of the fit to 0.4--8~keV. The {\it XMM-Newton}
spectra do not require any intrinsic absorption (the 90\% confidence
upper limit is $N_{\rm H}~<$~1.6$\times$10$^{20}$~cm$^{-2}$) and the
modest negative residuals in Figure 3a disappear in Figure 3d (top)
when the photon index is adjusted to
$\Gamma$~=~1.37$_{-0.05}^{+0.07}$. Applying this procedure to the {\it
Chandra} data does indeed indicate the presence of intrinsic
absorption, but the negative residuals in the {\it Chandra} spectrum
of Figure 3b are accomodated primarily through an extreme flattening
of the photon index to $\Gamma$~=~1.13$_{-0.09}^{+0.10}$. This model
is physically implausible in light of the inconsistency with the
best-fit high-energy photon index, and it is also not a particularly
good fit (${\chi}^{2}$/$\nu$ = 85/76). A lower limit to the intrinsic
absorption indicated by the {\it Chandra} spectrum may be obtained by
constraining $\Gamma$ to lie within the 90\% errors from the
high-energy fit; the resulting column density is $N_{\rm
H}~<$~9.29$\times$10$^{20}$~cm$^{-2}$, with $\Gamma$ fixed at
1.26. The residuals for this fit (shown in Figure 3d, middle) suggest
that further refinements to the absorption model are required.

Complex absorption is common in BAL quasars, of both the radio-quiet
(e.g., Gallagher et al.~2002) and the radio-loud (Brotherton et
al.~2005) types, and thus it is perhaps not surprising that the simple
intrinsic-absorption model is insufficient to fit the {\it Chandra}
spectrum for PG~1004+130. A partial-covering absorber model gives a
better representation (${\chi}^{2}$/$\nu$ decreases from 85/76 to
79/75, an improvement with an $F$-test probability of only 0.02 of
occuring by chance; see Figure 3d, bottom), with parameters $N_{\rm
H}$~=~1.20$_{-0.84}^{+0.83}\times$10$^{22}$ cm$^{-2}$,
$f_{c}$~=~0.49$_{-0.26}^{+0.14}$, and
$\Gamma$~=~1.37$_{-0.22}^{+0.18}$. Here $f_{c}$ is the fraction of the
source emission that passes through an intrinsic redshifted absorber
with column density given by the fitted $N_{\rm H}$, while the
remaining 1$-f_{c}$ of the source emission experiences only Galactic
absorption. Given our photon statistics, we cannot constrain the
nature of the absorption complexity in detail, but the physical
significance of this result is discussed further in $\S$5.

While variable absorption appears to be required in PG~1004+130, a
brief consideration of possible alternative explanations for the
discrepancy between the {\it XMM-Newton} and {\it Chandra} soft-band
spectra is warranted. The small-scale jet seen in the {\it Chandra}
image and unresolved by {\it XMM-Newton} is much too weak to explain
the differences in the soft X-ray spectra: when the larger {\it
XMM-Newton} extraction region, which includes the jet, is used to
extract a {\it Chandra} spectrum, the resulting parameters for the
partial-covering absorber model are similar to those found above, with
$N_{\rm H}$~=~0.95$_{-0.97}^{+0.81}\times$10$^{22}$ cm$^{-2}$,
$f_{c}$~=~0.42$_{-0.28}^{+0.17}$, and
$\Gamma$~=~1.30$_{-0.21}^{+0.20}$. Another possibility is that the
change in the low-energy X-ray spectrum may be due to variable
soft-band emission rather than variable absorption. It has been
suggested that the ``soft-excess'' in FR II objects may be related to
jet emission (e.g., Evans et al.~2006), and so perhaps a variable
unresolved jet might explain the discrepancy between the {\it
XMM-Newton} and {\it Chandra} spectra. The {\it XMM-Newton} spectra
can also be adequately fitted with a partial-covering absorber with
parameters fixed (except for normalization) to the best-fit {\it
Chandra} values, and an additional ``unresolved jet'' power-law
component with a likely photon index fixed at $\Gamma$ = 1.8. The
``unresolved jet'' would then have had to decrease in brightness by a
factor of $\sim$15 by the {\it Chandra} observation; flaring on that
order has been observed in the inner jet knot of M87 (Harris et
al.~2006), so it is possible (albeit somewhat contrived) that the {\it
XMM-Newton} observation occurred during a flaring episode. However,
the ``unresolved jet'' must contribute approximately one-fourth of the
total {\it XMM-Newton} 2--8 keV flux to avoid worsening the fit, and
as the {\it XMM-Newton} hard-band flux is already observed to be lower
than that measured by {\it Chandra}, additional variability of the
nuclear X-ray emission would be required. The coincidental brightening
of the ``unresolved jet'' during the presumed flare episode to the
level necessary to mimic the unbroken, continuous power-law seen in
the {\it XMM-Newton} spectra further suggests that variable absorption
is a more logical explanation than such conspiratorial combinations of
variable emission components. Note that disfavoring an additional
variable ``unresolved jet'' component does not impact interpretation
of the underlying origin of the entire 0.5--8 keV nuclear X-ray
emission, a topic considered further in $\S$5.

The 0.5--2~keV observed fluxes for the best-fit models as described
above are 1.07$\times$10$^{-13}$ erg~cm$^{-2}$~s$^{-1}$ for {\it
XMM-Newton} and 9.35$\times$10$^{-14}$ erg~cm$^{-2}$~s$^{-1}$ for {\it
Chandra}, with 2--8~keV fluxes of 2.60$\times$10$^{-13}$
erg~cm$^{-2}$~s$^{-1}$ and 3.56$\times$10$^{-13}$
erg~cm$^{-2}$~s$^{-1}$. The observed 1980 {\it Einstein}
\hbox{0.5--4.5}~keV soft-band limit would predict (for a $\Gamma$ =
1.5 power-law model with Galactic absorption) 0.5--2~keV and 2--8~keV
fluxes of 6.7$\times$10$^{-14}$ erg~cm$^{-2}$~s$^{-1}$ and
1.5$\times$10$^{-13}$ erg~cm$^{-2}$~s$^{-1}$, respectively, with a
greater uncertainly applying to the high-energy extrapolation; both
the {\it XMM-Newton} and the {\it Chandra} soft and hard-band fluxes
are higher than suggested by the {\it Einstein} non-detection. The
{\it Einstein} non-detection cannot be explained by simply increasing
the covering factor while holding the column density fixed to the
value measured in the {\it Chandra} spectrum, as even with $f_{c}$=1
the predicted 0.5--4~keV flux is higher than the limit measured by
{\it Einstein}. If the {\it Einstein} non-detection results from
variable absorption (rather than emission) then $N_{\rm H}$ must have
been a factor of $\sim$3 higher (for $f_{c}$=1) to account for the
lack of soft-band X-ray flux relative to the {\it Chandra}
observation. The unabsorbed 0.5--2~keV (rest-frame 0.6--2.5~keV)
luminosities are 2.08$\times$10$^{43}$ erg~s$^{-1}$ for {\it
XMM-Newton} and 2.98$\times$10$^{43}$ erg~s$^{-1}$ for {\it Chandra},
with 2--8~keV (rest-frame 2.5--9.9~keV) luminosities of
4.52$\times$10$^{43}$ erg~s$^{-1}$ and 6.44$\times$10$^{43}$
erg~s$^{-1}$.

There are sufficient counts to fit the {\it Chandra} X-ray jet
spectrum with a simple power-law model (assuming fixed Galactic
absorption), although the small number of counts necessitates use of
the XSPEC $C$-statistic rather than ${\chi}^{2}$. The best-fit photon
index is $\Gamma$~=~1.71$^{+0.51}_{-0.47}$. A power-law model is
almost universally appropriate for the \hbox{X-ray} spectra of jets
(e.g., Worrall \& Birkinshaw~2006), but with the limited counts
available here other possibilities cannot be excluded. For example, a
themal bremsstrahlung model also yields an acceptable fit (the
$C$-statistic value is essentially unchanged) with $kT\sim$5~keV,
although the temperature is poorly constrained. The 0.5--2~keV and
2--8 keV fluxes associated with the power-law model are 1.8 and
3.1$\times$10$^{-15}$ erg~cm$^{-2}$~s$^{-1}$, respectively,
corresponding to unabsorbed luminosities of 3.6 and
5.3$\times$10$^{41}$ erg~s$^{-1}$ (quoted for isotropic, unbeamed
emission; see $\S$4.2).

\section{Multiwavelength Properties}

\subsection{The Nucleus}

The optical/UV-to-X-ray spectral slope, ${\alpha}_{\rm ox}$, describes
the ratio of rest-frame luminosity density at 2500 {\AA} to that at
2~keV as ${\alpha}_{\rm ox}~=~0.384{\times}\log{(l_{\rm x}/l_{\rm
uv})}$; we take $l_{\rm uv}$ and $l_{\rm x}$ to have units of
erg~s$^{-1}$~Hz$^{-1}$ throughout. The optical/UV and X-ray
luminosities of radio-quiet quasars (RQQs) are correlated ($l_{\rm x}
\propto l_{\rm uv}^{\alpha}$), although not directly proportional
(e.g., Strateva et al.~2005; Steffen et al.~2006). This relationship
likely reflects a connection between optical/UV emission from the
inner accretion disk and X-ray emission arising from Compton
upscattering of disk photons in a hot corona. In RLQs this picture is
complicated by additional jet emission, and in both RQQs and RLQs
absorption can significantly depress X-ray flux. The ratio of
optical/UV to X-ray luminosity increases with increasing optical/UV
luminosity, which may be expressed as an anti-correlation between
${\alpha}_{\rm ox}$ and $l_{\rm uv}$; Steffen et al.~(2006) find
${\alpha}_{\rm ox}$~=~$-$0.139${\times}\log{l_{\rm uv}}$+2.680 for
their sample of unabsorbed RQQs. We compare the ${\alpha}_{\rm ox}$
value for PG~1004+130 with those of other quasars with $M_{B}<-$23 and
$z<$~0.5 in the Palomar-Green (PG; Schmidt \& Green~1983) survey.

We make use of data from Steffen et al.~(2006) for the PG
${\alpha}_{\rm ox}$ values. They determined monochromatic UV
luminosities by extrapolating known 3000 {\AA} values to 2500 {\AA}
assuming ${\alpha}_{\rm o}$~=~$-$0.5, and used {\it ROSAT} pointed and
All-Sky Survey PSPC count rates with an assumed $\Gamma$~=~2 power-law
to calculate monochromatic X-ray luminosities. The parameterization of
the luminosity dependence of ${\alpha}_{\rm ox}$ for RLQs is not as
accurately known as it is for RQQs. \citet{1987ApJ...313..596W} find
that RLQs with flat radio spectra also have somewhat flatter
optical/UV-to-X-ray spectra than steep-spectrum RLQs (which they
attribute to stronger jet X-ray emission in the flat-spectrum
sources), but both flat-spectrum and steep-spectrum RLQs have $l_{\rm
x} \propto l_{\rm uv}^{\alpha}$ correlations with slopes consistent
(within the 90\% error range) with those of RQQs. We therefore remove
the luminosity dependence of ${\alpha}_{\rm ox}$ using the Steffen et
al.~(2006) relation for RQQs discussed previously, which we find also
orders RLQs adequately for purposes of comparison. The resulting
${\alpha}_{\rm ox}~-~{\alpha}_{\rm ox}(l_{\rm uv})$ histograms are
shown in Figure 4a. While there are some quasars in the negative
${\alpha}_{\rm ox}$ tail of the RQQ distribution that lack evidence of
intrinsic UV or X-ray absorption, those quasars with strong UV
absorption all have anomalously steep optical/UV-to-X-ray spectral
slopes. The two RLQs with confirmed intrinsic absorption (PG 1309+355
and 2251+113 both show UV and X-ray absorption) similarly have more
negative values of luminosity-corrected ${\alpha}_{\rm ox}$ than most
of the unabsorbed RLQs.

The monochromatic UV luminosity for PG~1004+130 was determined from
the {\it XMM-Newton} OM fluxes and SDSS spectroscopy, as the dates of
these measurements are most nearly coincident with those of the X-ray
observations. The SDSS spectrum was scaled by SDSS photometric
measurements to correct for fiber inefficiencies. Extrapolation to
rest-frame 2500 {\AA}~was performed by renormalizing a standard
\citep{1994ApJS...95....1E} RLQ SED to fit the dereddened
\citep{1989ApJ...345..245C} optical/UV data; the scaled RLQ SED was
matched to the OM UVW1 and UVM1 fluxes and is then only slightly below
the continuum of the SDSS spectrum, giving a luminosity density of
$\log{l_{\rm uv}}~$=~30.51. The monochromatic X-ray luminosity was
calculated from the {\it Chandra} data for two different best-fit
models (see $\S$3 for details), the first a simple power-law fit over
the observed \hbox{2--8}~keV band (yielding a ``hard'' ${\alpha}_{\rm
ox}$~=~$-$1.83 determined by the high-energy X-ray spectrum) and the
second a partial-covering neutral absorption model fit over the
observed 0.5--8~keV band (yielding a ``soft'' ${\alpha}_{\rm
ox}$~=~$-$1.88 primarily influenced by the low-energy X-ray
spectrum). The model flux densities were measured at rest-frame 2~keV
(observed 1.6~keV) and converted to bandpass-corrected luminosity
densities. The results of this analysis are plotted on Figure 3, along
with the {\it Einstein} ${\alpha}_{\rm ox}<-$2.01 limit from
\citet{1984ApJ...280...91E}. We note that our $l_{\rm uv}$ is
$\approx$ 2.5 times less than that used by \citet{1984ApJ...280...91E}
to determine ${\alpha}_{\rm ox}$, and the {\it Einstein} X-ray flux
limit with our $l_{\rm uv}$ measurement would give ${\alpha}_{\rm
ox}<-$1.97. The 1973--1990 photographic monitoring of Smith et
al.~(1993) indicates that PG~1004+130 fluctuates in optical brightness
by $\approx$~0.5 mag on timescales of 6--10 yr. More recently, Garcia
et al.~(1999; 2006, private communication) found that PG~1004+130
brightened and then dimmed over a magnitude range of $V$ = 14.7--15.2
from 1993--1999 (the SDSS 2003 photometry corresponds to $V$ = 15.3),
and Stalin et al.~(2004) found that PG~1004+130 dimmed in $R$ by 0.09
magnitudes from 1999 March to 2000 April. The {\it Einstein} vs. {\it
Chandra} discrepancy in ${\alpha}_{\rm ox}$ values arises from a
combination of measurement uncertainties and genuine variability at
both optical/UV and X-ray frequencies. The optical/UV-to-X-ray
spectral slope is indeed steeper for PG~1004+130 than for other PG
RLQs, even if calculated from the hard-band emission, and it is
steeper than almost all of the non-absorbed RQQs as well. To
differentiate conclusively between excess ``big blue bump'' emission
(see Elvis \& Fabbiano 1984) and X-ray weakness as the cause of the
low value of ${\alpha}_{\rm ox}$, it is helpful to compare the
broad-band SED of PG~1004+130 with those of other quasars.

The SED of PG~1004+130 presented in Figure 4b was constructed with
radio and IR fluxes, optical and UV spectra and photometry, and the
X-ray best-fit models. There are two sets of radio measurements shown:
the Parkes data include the extended radio emission, while the {\it
VLA} data are for the nucleus alone. The optical (SDSS) and UV ({\it
HST} GHRS and {\it IUE\/}) spectra were smoothed to reduce noise, and
the geocoronal Ly$\alpha$ region has been excluded. The SDSS spectrum
was scaled to match SDSS photometric measurements, as described
previously. Data were corrected for Galactic extinction (with
$E(B-V)$~=~0.038~mag) following \citet{1989ApJ...345..245C}. We have
included the best-fit {\it XMM-Newton} power-law model and the {\it
Chandra} partial-covering absorber model with parameters as given in
$\S$3 (in both cases correcting for Galactic absorption), as well as
the 2~keV flux density corresponding to the 0.5--4.5~keV {\it
Einstein} limit. Bandpass-corrected luminosity densities at
radio-to-X-ray rest-frame frequencies for PG~1004+130 are given in
Table 1.

Comparing the standard quasar SEDs compiled by
\citet{1994ApJS...95....1E} to that of PG~1004+130 in Figure 4b, there
is excellent agreement in the shapes of the SEDs at
radio-to-optical/UV frequencies. There does not appear to be evidence
for enhanced UV emission (relative to the radio, IR, and optical
data), but PG~1004+130 is distinctly X-ray weak relative to RLQs with
comparable optical/UV luminosities. Removing the intrinsic absorption
apparent in the {\it Chandra} spectrum partially accounts for the
weakness of the X-ray emission, but the unabsorbed {\it Chandra} and
{\it XMM-Newton} power-law spectra remain below the standard RLQ X-ray
emission. Some of this apparent X-ray weakness may be a consequence of
the chosen method of comparison, as the Elvis et al.~(1994) composite
SEDs are biased toward X-ray bright objects due to selection criteria
requiring an {\it Einstein} detection. Indeed, the average
${\alpha}_{\rm ox}$ of the RLQs used to contruct the standard SED
plotted in Figure 4b is $\langle{\alpha}_{\rm ox}\rangle$~=~$-$1.31,
corresponding to relatively greater X-ray luminosities than the
$\langle{\alpha}_{\rm ox}\rangle$~=~$-$1.54 average of the PG RLQs.
However, the ``hard'' ${\alpha}_{\rm ox}$~=~$-$1.83 for PG~1004+130 is
still 1.2$\sigma$ below the average ${\alpha}_{\rm ox}$ of the PG
RLQs; the corresponding ratio by which PG~1004+130 is X-ray weak
relative to other PG RLQs is $\sim$5.4 (for similar optical/UV
luminosities).

The black-hole mass for PG~1004+130 is found by Vestergaard \&
Peterson~(2006) to be
1.87$^{+0.40}_{-0.40}~\times$~10$^{9}$~M$_{\odot}$ (from the FWHM of
the H$\beta$ line and the monochromatic optical luminosity at
5100~\AA), while Falomo et al.~(2003) estimate a mass of
1.35~$\times$~10$^{9}$~M$_{\odot}$ (from the $M_{\rm BH}$--$L_{\rm
bulge}$ relation). The black-hole masses calculated for PG~1004+130
using these different methods are in general agreement and correspond
to an Eddington luminosity of $\sim$2~$\times$~10$^{47}$
erg~s$^{-1}$. We measure the bolometric luminosity for PG~1004+130 by
integrating the scaled standard SED up to X-ray frequencies, then
integrating the best-fit power-law model from rest-frame 0.5--10~keV,
and obtain $L_{\rm Bol}$ = 2.0~$\times$~10$^{46}$ erg~s$^{-1}$. The
observed X-ray emission may not be representative of the true X-ray
power of the source (see $\S$5); if instead the scaled standard SED is
integrated to 10~keV, the bolometric luminosity is slightly higher,
$L_{\rm Bol}$ = 2.3~$\times$10$^{46}$ erg~s$^{-1}$. Both calculations
suggest that PG~1004+130 is radiating at $L_{\rm
Bol}~\sim$~10$^{-1}$~L$_{\rm Edd}$.

\subsection{The Jet}

The SE radio jet is made up of a string of emission peaks (knots),
presumably indicating distinct shock sites where particle acceleration
takes place. The spectral index ${\alpha}_{\rm r}$ steepens downstream
of each knot (see Figure 1c), perhaps reflecting an evolution in the
underlying electron distribution. The observed-frame lifetimes for
synchrotron cooling for plausible magnetic-field strengths are quite
long, corresponding to scales orders of magnitude longer than these
projected distances, so if spectral aging is the dominant factor
behind this effect the electrons must remain trapped within the shock
region for long periods before diffusing downstream. The X-ray
emission begins upstream of the first bright radio knot, and there
does not appear to be any jet-associated optical emission. Radio
fluxes were extracted from the region encompassing the first knot,
overlapping with the end of the X-ray jet extraction region indicated
in Figure 1a and extending a short distance beyond it. A 5$\sigma$
optical upper limit was determined from the noise in the {\it HST}
image within the X-ray jet extraction region. The 2~keV ${\nu}S_{\nu}$
taken from the power-law fit to the jet X-ray spectrum and the error
bars on ${\alpha}_{\rm x}$ are plotted in Figure~5 along with the {\it
HST} limit and the {\it VLA} fluxes. The power-law fit for the jet is
consistent with the $\Gamma~\sim$~1.5 photon indices found by Sambruna
et al.~(2004) for the brightest X-ray knots in their {\it Chandra} and
{\it HST} survey of core-dominated FR~II quasars with known radio
jets; however, the X-ray spectra of prominent knots in FR~I jets are
generally steeper, with $\Gamma~\sim$~2.3 (e.g., 3C~66B: Hardcastle et
al.~2001; 3C~31: Hardcastle et al.~2002; M~87: Marshall et al.~2002;
Cen~A: Hardcastle et al.~2003; B2~0755+37: Parma et al.~2003). The
${\alpha}_{\rm rx}$ = $-$0.87 value for PG~1004+130 is similar to
those seen for both FR~I (FR~I jet references as above) and FR~II
(Sambruna et al.~2004; Marshall et al.~2005) jets. The PG~1004+130 jet
optical limit falls well below the power law connecting the radio and
X-ray data, ruling out simple single-component synchrotron models. The
broad-band spectral indices are constrained to be ${\alpha}_{\rm
ro}<-$1.1, ${\alpha}_{\rm ox}>-$0.46, and consequently ${\alpha}_{\rm
ro}/{\alpha}_{\rm ox}>$~2.3. These values are similar to those found
for FR~II jets (Sambruna et al.~2004) but are inconsistent with those
of FR~I jets, which tend to have ${\alpha}_{\rm ro}\sim-$0.7,
${\alpha}_{\rm ox}\sim-$1.2, and thus ${\alpha}_{\rm ro}/{\alpha}_{\rm
ox}\sim$0.6 (FR~I jet references as above). Despite its standard FR~I
radio structure, the SE PG~1004+130 jet shares many of the
characteristics of well-known FR~II jets.

While the agreement between the multiwavelength characteristics of the
SE FR~I jet and powerful FR~II jets initially appears somewhat
surprising, this result might have been anticipated based on the radio
luminosity and hybrid morphology of PG~1004+130. The radio luminosity
density at rest-frame 178 MHz of PG~1004+130 is
$\sim$~7.1~$\times$~10$^{33}$ erg~s$^{-1}$~Hz$^{-1}$ (from the flux
measurement of Wright \& Otrupcek 1990), more than an order of
magnitude above the 2~$\times$~10$^{32}$ erg~s$^{-1}$~Hz$^{-1}$
luminosity density found by Fanaroff \& Riley (1974) to divide
empirically the lower power FR~I population from the higher power
FR~II sources. More recently, the radio power separating FR~I from
FR~II sources has been observed to be an increasing function of the
host galaxy optical luminosity (e.g., Ledlow \& Owen 1996), and the
optical magnitude ($M_{\rm R}~=~-$24.26; Falomo et al.~2003) of its
elliptical host \citep{1997ApJ...479..642B} places PG~1004+130
somewhat closer to the observed transition line.\footnotemark[3]
\footnotetext[3]{As noted by Scarpa \& Urry~(2001), the underlying
radio and optical luminosity functions lead naturally to an
anti-correlation of radio power with host luminosity, and hence host
luminosity need not be physically indicative of radio
characteristics.} Further, the NW lobe of PG~1004+130 has standard
edge-brightened FR~II structure. If hybrid morphology sources are
reflective of dissimilar environments rather than dissimilar jets, as
suggested by Gopal-Krishna \& Wiita (2000), then the SE FR~I jet
should be as intrinsically powerful as the NW FR~II jet.

We have applied various models to the multiwavelength jet emission
with the goal of determining the most plausible origin for the X-ray
emission. The results of this analysis are described below and
representative models are shown in Figure 5. The radio-to-optical
emission in jets is well-established as synchrotron radiation, with
the principle observational support coming from polarization
measurements. The radio data and optical limit allow determination of
the magnetic field, assuming equipartition. Based on the {\it VLA}
images we estimate the emission region to be roughly circular with a
radius of around 1.75$''$, corresponding to a spherical volume of
3.6$\times$10$^{67}$~cm$^{3}$. As is standard practice, we assume a
power-law electron energy distribution, with index $p$=2 to match the
spectral slope of the {\it VLA} radio data (see Figure 1c). The low
energy cutoff for the electron spectrum is observationally
unconstrained, and we choose ${\gamma}_{\rm min}$=50, similar to
values typically assumed for FR~II jets. The high energy cutoff is
limited by the {\it HST} non-detection, and we use ${\gamma}_{\rm
max}$=10$^{6}$. In the case of PG~1004+130, the large angular size,
the high lobe-to-core ratio, and the optical-to-radio core luminosity
ratio suggest the jet is inclined to the line-of-sight by
$\theta~\simgt$~45$^{\circ}$, which would limit the allowed beaming to
$\delta~\simlt$~1.4. We consider here the $\delta$=1 case. We use
standard synchrotron formulae (e.g., Worrall \& Birkinshaw 2006) for a
single-injection model with pitch-angle isotropization (Jaffe \&
Perola 1973). A continuous-injection model, in which the spectral
slope steepens by 0.5 above a critical frequency (e.g., Carilli et
al.~1991), would still require an exponential cutoff at frequencies
below the optical limit, but a Kardashev-Pacholczyk (Kardashev 1962;
Pacholczyk 1970) model with no pitch-angle scattering would allow a
synchrotron cutoff at higher frequencies. The precise slope of the
synchrotron spectrum above the turnover frequency does not greatly
affect the X-ray emission, so a Kardashev-Pacholczyk or
continuous-injection model would lead to similar qualitative
conclusions. We derive a magnetic field strength $B_{1}$ = 14~$\mu$G,
and note that in general $B_{eq}$ = $B_{1}$/$\delta$ (e.g., Harris \&
Krawczynski 2002).

A natural explanation for the origin of the X-ray emission would be
Compton upscattering by the synchrotron electrons. We consider two
sources of seed photons: the synchrotron radiation itself (SSC; e.g.,
Hardcastle et al.~1997) or the cosmic microwave background (IC/CMB;
e.g. Tavecchio et al.~2000; Celotti et al.~2001). At the distance of
the X-ray jet, the photon flux from the AGN or from the host galaxy is
comparatively insignificant. The X-ray emission expected from the SSC
process with the above parameters is more than three orders of
magnitude less than the observed X-ray flux. The X-ray SSC emission
increases relative to the synchrotron emission if the magnetic-field
strength decreases, but radically sub-equipartition fields are
required to attribute the observed X-ray flux to SSC emission. A
representative model of this type is included in Figure 5. Both the
high and low electron-energy cutoffs have been adjusted to accommodate
the optical limit (10$^{3.2}~<~\gamma~<~$10$^{7.5}$) but given the
$p$~=~2 power-law distribution and the consistent span of 4.3 decades
in energy the equipartion magnetic field remains 14 $\mu$G. The actual
magnetic field required to fit the X-ray flux is then 0.021 $\mu$G,
several hundred times less than the equipartion value. IC/CMB is often
put forward as an explanation for the concave SEDs of powerful FR~II
jets, and this process is particularly efficient for high-redshift,
relativistic jets inclined close to the line of sight. Unlike SSC,
which is actually depressed by beaming, X-ray IC/CMB emission is
boosted by an additional factor of 1+$\alpha$ relative to synchrotron
emission (Dermer 1995). In the case of PG~1004+130, the expected
IC/CMB emission is still several hundred times less than the observed
X-ray emission; the enhanced beaming required to match the X-ray flux,
$\delta$=3.0, would place an upper limit (where $\delta$ = $\Gamma$)
on the jet angle of $\theta < \arcsin{{\delta}^{-1}} < $19$^{\circ}$,
smaller than the $\theta~\simgt$~45$^{\circ}$ suggested from the
optical and radio luminosities as well as the radio morphology (see
$\S$1). Further, the lifetime of the low-energy electrons ($\gamma
\sim$ 10$^{2}$) responsible for the X-ray emission in IC/CMB models
greatly surpasses the length of the jet, and so X-ray emission would
be expected to persist along the jet (absent deceleration; e.g.,
Georganopoulos \& Kazanas~2004). Despite the similarities in
PG~1004+130 to the SEDs and X-ray spectra of the core-dominated,
highly beamed FR~II jets that dominate the surveys of both Sambruna et
al.~(2004) and Marshall et al.~(2005), the IC/CMB model commonly
applied to such FR~II jets does not appear to be appropriate for
PG~1004+130. For both SSC and IC/CMB models, the location of the X-ray
emission largely upstream of the parent synchrotron electrons is
difficult to explain and the predicted X-ray spectral slope is flatter
than observed. We consider it unlikely that the X-ray emission arises
from either SSC or IC/CMB emission.

If the X-ray jet emission does not arise as a consequence of the
low-energy synchrotron radiation (as in the SSC and IC/CMB cases),
then various other emission mechanisms may be considered, such as
thermal bremsstahlung or synchrotron emission from a secondary
population of high-energy electrons (e.g., Atoyan \& Dermer 2004). The
offset of the X-ray and radio emission is perhaps more easily
accomodated by such models. The X-ray emission occupies a rectangular
region of approximately 5$''$ by 2.4$''$; for an edge-on cylinder this
again corresponds to a volume of 3.6$\times$10$^{67}$~cm$^{3}$. The
best-fit temperature for a bremsstrahlung model is $kT~\sim~$5~keV,
but this is only poorly constrained. Neglecting line emission (which
contributes significantly to the soft X-ray emission at lower
temperatures), a gas cloud with an average ion charge of $Z\sim$1
would be required to have a density of $n~=~$0.05~cm$^{-3}$ to account
for the observed X-ray flux. This corresponds to a total mass of
1.5$\times$10$^{9}$~M$_{\odot}$ and an ideal-gas pressure of
4.6$\times$10$^{-10}$ dynes cm$^{-2}$. Such a large quantity of
concentrated hot gas at so great a distance from the host galaxy seems
unlikely, and as this gas cloud would be overpressured with respect to
the surrounding IGM we would have to be observing it at a favorable
time before it dispersed. If instead the X-ray emission from
PG~1004+130 arises from a second synchrotron component generated by a
population of highly relativistic electrons, the low-energy cutoff
must be high enough to avoid over-predicting the optical flux. Taking
the magnetic field to be 14~$\mu$G and setting the electron energy
index to be $p$=2.4 as indicated by the X-ray photon index, this model
provides an acceptable explanation of the X-ray emission with
${\gamma}_{\rm min}$=1.6$\times$10$^{7}$, as shown in Figure 5. As the
lifetime of the X-ray synchrotron electrons is quite short (electrons
initially associated with 1 keV emission would have half lives of
$\sim$600 years), multiple acceleration or injection sites are
required along the extent of the X-ray jet. Electrons with these
injection parameters would lose sufficient energy within $\sim$30,000
years to produce fluxes at optical frequencies in excess of the
observed {\it HST} limit, suggesting that the energetic electrons
escape the shock region before cooling to energies below
$\gamma\sim$3$\times$10$^{6}$.

The diffuse emission extends 40--50$''$ (150--190 projected kpc) from
the nucleus to both the SE and the NW, tracing the path of the FR~I
radio jet and providing supporting evidence for the presence of the
hidden FR~II counterjet. There are insufficient counts for spectral
analysis, but with an assumed $\Gamma$=1.8 power law the 0.5--8 keV
X-ray flux of the SE component is 4.5$\times$10$^{-15}$
erg~cm$^{-2}$~s$^{-1}$, while that of the NW component is
4.2$\times$10$^{-15}$ erg~cm$^{-2}$~s$^{-1}$. Most radio jets are
one-sided, with the absence of a detectable counterjet generally
attributed to Doppler boosting and hence yielding a constraint upon
the line-of-sight angle. As the line-of-sight angle for PG~1004+130 is
likely $\theta~\simgt$~45$^{\circ}$, the jet/counterjet flux ratio for
twin jets is expected to be less than $\sim$15 (using $R_{\rm
J}~=~[(1-\beta\cos{\theta})/(1+\beta\cos{\theta})]^{\alpha-2}$ and
$\Gamma$ = ${\delta}_{\rm max}$ = 1/$\sin{\theta}$ = 1.4; e.g.,
Worrall \& Birkinshaw 2006). However, the observed ratio of radio
emission in the SE jet to that in the undetected NW counterjet exceeds
100, indicating that Doppler boosting alone cannot explain the lack of
a radio counterjet in PG~1004+130. The diffuse X-ray emission
precludes the possibility of an intrinsically one-sided jet, leading
us to hypothesize that the entraining environment is indeed less dense
to the NW, as suggested by the hybrid radio morphology.

If the diffuse X-ray emission is thermal radiation, then the required
gas density is $\sim$4.5$\times$10$^{-4}$~cm$^{-3}$ with a total mass
of 2.3$\times$10$^{11}$~M$_{\odot}$ and a pressure of
1.5$\times$10$^{-12}$ dynes cm$^{-2}$; while these parameters are not
as restrictive as those for the thermal jet models, similar concerns
apply. The mechanism for heating the gas along the jet is unclear, and
a non-thermal origin seems somewhat more plausible. SSC emission alone
is not a viable explanation for the diffuse X-ray emission, as the
absence of a detectable radio counterjet to the NW imposes a stringent
limit to the available synchrotron photon density above 1 GHz, and the
concurrent lack of energetic electrons makes boosting to X-ray
frequencies difficult. The diffuse X-ray emission could result from
unbeamed IC/CMB emission, and indeed its apparently smooth extent
along the entire jet suggests an association with long-lived
electrons. The paucity of radio emission in the NW counterjet region
does not greatly affect the IC/CMB X-ray yield, which is driven by
low-energy electrons associated with sub-GHz synchrotron
radiation. Diffuse X-ray emission unaccompanied by detectable radio
emission has been observed in a handful of additional sources and can
be successfully attributed to IC/CMB processes in those cases as well
(e.g., Siemiginowska et al.~2002; Schwartz et al.~2005). The {\it VLA}
C band limit on radio emission in the NW counterjet allows for
considerable leeway in determing the high-energy cutoff and magnetic
field required to generate sufficient diffuse X-ray emission via the
IC/CMB process. For ${\gamma}_{min}$ = 50 and ${\gamma}_{max}$ =
10$^{3.5}$, the magnetic field is required to be lower ($\sim$2
$\mu$G) than in the jet region, with equipartition electron densities
$\sim$30 times less than in the knots.

\section{Results and Discussion}

The primary results from our analysis of the first X-ray detections
and spectra of PG~1004+130 are the following:

1. Variable complex absorption: Although the {\it XMM-Newton}
   observation of PG~1004+130 shows only minimal intrinsic absorption,
   the {\it Chandra} spectrum reveals significant soft X-ray
   absorption that cannot be modeled by a simple redshifted neutral
   absorber. The best-fit parameters for the preferred
   partial-covering model are $N_{\rm
   H}$~=~1.20$_{-0.84}^{+0.83}\times$10$^{22}$ cm$^{-2}$,
   $f_{c}$~=~0.49$_{-0.26}^{+0.14}$, and
   $\Gamma$~=~1.37$_{-0.22}^{+0.18}$. The 2--8~keV {\it XMM-Newton}
   flux is 23\% lower than the 2--8~keV {\it Chandra} flux.

2. X-ray weakness: PG~1004+130 has the lowest luminosity-corrected
   value of ${\alpha}_{\rm ox}$ among the PG RLQs, and after
   correcting for intrinsic absorption it is $\sim$5.4 times weaker in
   X-rays than the other PG RLQs, when normalizing to similar
   optical/UV luminosities. Examination of the SED of PG~1004+130
   confirms that the anomalous value of ${\alpha}_{\rm ox}$ is due to
   X-ray weakness rather than optical/UV brightness.

3. X-ray jet: The {\it Chandra} image reveals an X-ray counterpart to
   the SE radio FR~I jet. The jet is undetected by {\it HST}, ruling
   out simple single-component synchrotron emission models, and both
   the X-ray spectrum and the multiwavelength SED shape are similar to
   those observed for prominent knots in FR II jets. Diffuse X-ray
   emission is observed along the path of the jet and counterjet.

Complex X-ray absorption is common in radio-quiet BAL quasars (e.g.,
Gallagher et al.~2002, 2006) and may be inferred by X-ray weakness
coupled with relatively soft X-ray spectra for radio-loud BAL quasars
as well (Brotherton et al.~2005). There is precedent for variable
X-ray absorption such as that seen in PG~1004+130; Gallagher et
al.~(2004) found that PG~2112+059 showed a factor of $\sim$7 increase
in intrinsic $N_{\rm H}$ over three years (483 rest-frame days). The
more absorbed {\it Chandra} spectrum for PG~2112+059 required either
an ionized or partially covering absorber to fit the flat soft X-ray
region and also revealed Fe K$\alpha$ emission undetectable against
the higher continuum of the earlier {\it ASCA} observation, with a
rest-frame equivalent width of 1050$^{+520}_{-471}$~eV. The best-fit
$N_{\rm H}$ for PG~1004+130 is relatively low compared with that of
most radio-quiet BAL quasars, as is the covering fraction (cf. Green
et al. 2001). Together these results suggest that the nature of the
X-ray absorber is broadly similar in radio-quiet and radio-loud BAL
quasars, but that the column density associated with the observed
X-ray absorption is lower in radio-loud BAL quasars, somewhat
analogous to the trend for UV absorption to be stronger in radio-quiet
BAL quasars than in radio-loud BAL quasars
\citep{2000ApJ...538...72B}.

Although notably X-ray weak relative to non-BAL RLQs, PG~1004+130 is
not as X-ray weak as the BAL RLQs examined by Brotherton et
al.~(2005), and it is possible that PG~1004+130 is simply an
intrinsically X-ray weak RLQ that also shows variable
absorption. Alternative hypotheses as to the cause of the X-ray
weakness of PG~1004+130 are constrained by the characteristics of the
{\it XMM-Newton} and {\it Chandra} spectra. The X-ray weakness cannot
be attributed solely to attenuation by simple intrinsic neutral
absorption, as the required high column densities would result in
X-ray spectra significantly harder than observed. If much of the
intrinsic X-ray emission is hidden by heavy absorption of
significantly higher column density ($N_{\rm
H}>~$5$\times$10$^{23}$~cm$^{-2}$) than that indicated by the flat
spectral shape, the observed X-ray spectrum could result from photons
leaking through ``holes'' in the absorber or scattering off a
``torus'' or electron-cloud mirror into the line of sight. However,
the absence of strong Fe K$\alpha$ emission is somewhat surprising if
the latter scenario is correct, and PG~1004+130 is actually brighter
in X-rays than expected for a reflection-dominated continuum. If the
nucleus is obscured for photon energies up to $\approx$~8~keV (or
extremely weak in X-rays), it is also possible that we are viewing
X-ray emission from a subparsec-scale jet; this may provide a natural
explanation for the absence of prominent Fe K$\alpha$ emission, as
well as the observed lack of excess low-energy emission that
distinguishes the X-ray spectra of PG~1004+130 from those of other
quasars.

Motivated by the observed correlation between radio and optical
luminosity for narrow-line radio galaxies, Chiaberge et al.~(2000)
suggested a synchrotron origin for the nuclear optical emission from
these sources. This correlation has been extended to the X-ray band by
Evans et al.~(2006), who argue that FR I RLQs derive a significant
fraction of their (generally unabsorbed) nuclear X-ray emission from
an unresolved jet, while FR II RLQs are dominated by (absorbed)
accretion-powered X-ray emission but also contain a jet spectral
component. In the particular case of PG~1004+130, the complexity and
intermediate column density of the absorption in the observed {\it
Chandra} X-ray spectrum could plausibly be attributed to viewing the
jet through progressively diminishing BAL-wind column densities with
increasing distance from the nucleus. X-ray spectra with improved
photon statistics and resolution are required to constrain better the
physical nature of the nuclear X-ray emission, or it may also be
possible to detect direct X-rays in the $\approx$ 8--200~keV band if
the putative absorption does not exceed $N_{\rm
H}~\simeq~$2$\times$10$^{24}$~cm$^{-2}$ (e.g., Matt~2002). Detection
of rapid X-ray variability would restrict the size of the emission
region and limit the degree to which reprocessed radiation could
contribute to the observed continuum, but neither the {\it XMM-Newton}
nor the {\it Chandra} observation shows such variability.

Identification of the X-ray emission mechanism in the PG~1004+130 jet
would be aided by an optical detection of the jet and determination of
the frequency and nature of the break in the SED between radio and
X-ray wavelengths. However, we consider it unlikely that the X-ray jet
is dominated by IC/CMB emission, and instead favor the X-ray emission
arising from a second synchrotron component. As with other X-ray jets
(e.g., Marshall et al.~2002; Sambruna et al.~2004), the X-ray jet of
PG~1004+130 peaks in brightness upstream of the brightest radio
knot. This suggests that X-ray synchrotron emission may be more
prominent in the inner jet, either because of stronger magnetic fields
or a supply of highly energetic electrons (e.g., Sambruna et
al.~2004). The agreement in the multiwavelength properties of the
PG~1004+130 jet with those of other FR~II jets indicates that the FR~I
radio morphology is likely due to propagation into a dense environment
rather than intrinsically lower power, as suggested by
\citet{2000A&A...363..507G}. The orientation of the jet axis
demonstrates that concave jet SEDs can arise without substantial
beaming. The diffuse X-ray emission traces the path of the jet and
hidden counterjet and suggests that low-energy electrons inhabit the
entire length from nucleus to lobes, more consistent with continuous
jet emission from the central engine than with sporadic activity
(e.g., Stawarz et al.~2004).

\acknowledgments

We gratefully acknowledge the financial support of NASA grant SAO
SV4-74018 (GPG, Principal Investigator), NASA LTSA grant NAG5-13035
(BPM, WNB, DPS), {\it XMM-Newton} grant NAG5-13541 (BPM, WNB), the
{\it Spitzer} Fellowship Program, under award 1256317 (SCG), and NASA
grant GO-09432 from the Space Telescope Science Institute, which is
operated by the Association of Universities for Research in Astronomy,
Inc., under NASA contract NAS5-26555 (BJW). Ed Fomalont kindly reduced
and analyzed the {\it VLA} data used in this paper, and we appreciate
his generous assistance with this project. We thank George Chartas for
helpful discussions, and we thank an anonomous referee for many
constructive suggestions. This work includes observations obtained
with {\it XMM-Newton}, an ESA science mission with instruments and
contributions directly funded by ESA Member States and NASA. Funding
for the SDSS and SDSS-II has been provided by the Alfred P. Sloan
Foundation, the Participating Institutions, the National Science
Foundation, the U.S. Department of Energy, the National Aeronautics
and Space Administration, the Japanese Monbukagakusho, the Max Planck
Society, and the Higher Education Funding Council for England. The
SDSS Web Site is http://www.sdss.org/.

\clearpage

\begin{table}
\begin{center}
\caption{Multiwavelength Luminosity Data}
\begin{tabular}{cccll}
\tableline\tableline
Band & Frequency\tablenotemark{a} & Luminosity\tablenotemark{b} & Date\tablenotemark{c} & Source \\
\tableline
& 8.00 & 34.26 & \nodata & PKS 80 MHz \\[-1pt]
& 8.34 & 33.85 & \nodata & PKS 178 MHz \\[-1pt]
Radio\tablenotemark{d} & 8.70 & 33.58 & \nodata & PKS 408 MHz \\[-1pt]
& 9.24 & 33.26 (31.92) & \nodata & PKS (VLA) 1.4 GHz \\[-1pt]
& 9.53 & 33.05 & \nodata & PKS 2.7 GHz \\[-1pt]
& 9.79 & 32.77 (31.70) & \nodata & PKS (VLA) 5 GHz \\
\tableline
& 12.57 & 32.85 & 1983 & IRAS 100 $\mu$m \\[-1pt]
& 12.79 & 32.40 & 1983 & IRAS 60 $\mu$m \\[-1pt]
& 13.17 & 32.24 & 1983 & IRAS 25 $\mu$m \\[-1pt]
& 13.49 & 31.98 & 1983 & IRAS 12 $\mu$m \\[-1pt]
& 13.55 & 31.56 & 1986/01 & $N$ (BJW\tablenotemark{e}) \\[-1pt]
IR & 13.89 & 31.40 & 1986/01 & $M$ (BJW\tablenotemark{e}) \\[-1pt]
& 13.98 & 31.23 & 1986/01 & $L'$ (BJW\tablenotemark{e}) \\[-1pt]
& 14.03 & 31.17 & 1986/01 & $L$ (BJW\tablenotemark{e}) \\[-1pt]
& 14.23 & 30.83 & 2000/04/06 & 2MASS $K$ \\[-1pt]
& 14.35 & 30.69 & 2000/04/06 & 2MASS $H$ \\[-1pt]
& 14.47 & 30.67 & 2000/04/06 & 2MASS $J$ \\
\tableline
& 14.62 & 30.64 & 2003/01/28 & SDSS $z$ \\[-1pt]
& 14.70 & 30.66 & 2003/01/28 & SDSS $i$ \\[-1pt]
Optical & 14.78 & 30.61 & 2003/01/28 & SDSS $r$ \\[-1pt]
& 14.90 & 30.61 & 2003/01/28 & SDSS $g$ \\[-1pt]
& 15.02 & 30.61 & 2003/01/28 & SDSS $u$ \\
\tableline
& 15.11 & 30.48 & 2003/05/04 &  OM UVW1 \\[-1pt]
UV & 15.21 & 30.44 & 2003/05/04 &  OM UVM1 \\[-1pt]
& 15.24 & 30.38 & 2003/05/04 &  OM UVW2 \\[-1pt]
& 15.59 & 29.48 & 2000/12/17 &  FUSE    \\
\tableline
& 17.68 & 25.36 & 1980/05/09 & {\it Einstein} \\[-1pt]
X-ray& 17.68 & 25.59 & 2005/01/05 & {\it Chandra}  \\[-1pt]                    
& 17.68 & 25.56 & 2003/05/04 & {\it XMM-Newton} \\
\tableline
\end{tabular}

\tablenotetext{a}{{\rm L}og of rest-frame frequency in units of Hz.}
\tablenotetext{b}{{\rm L}og of bandpass-corrected luminosity density
in units of erg s$^{-1}$ Hz$^{-1}$, after correcting for Galactic
extinction.}  \tablenotetext{c}{Date is UT yyyy/mm/dd}
\tablenotetext{d}{Parkes catalog data from Wright \& Otrupcek~1990;
note that due to limited resolution the luminosity densities from the
Parkes survey quoted here are dominated by the extended
emission. Luminosities for the nucleus alone measured from {\it VLA}
data are given in parentheses.} \tablenotetext{e}{Data obtained by
author B.J.~Wills}

\end{center}
\end{table}
\clearpage

\begin{figure}
\epsscale{1.1}
\plottwo{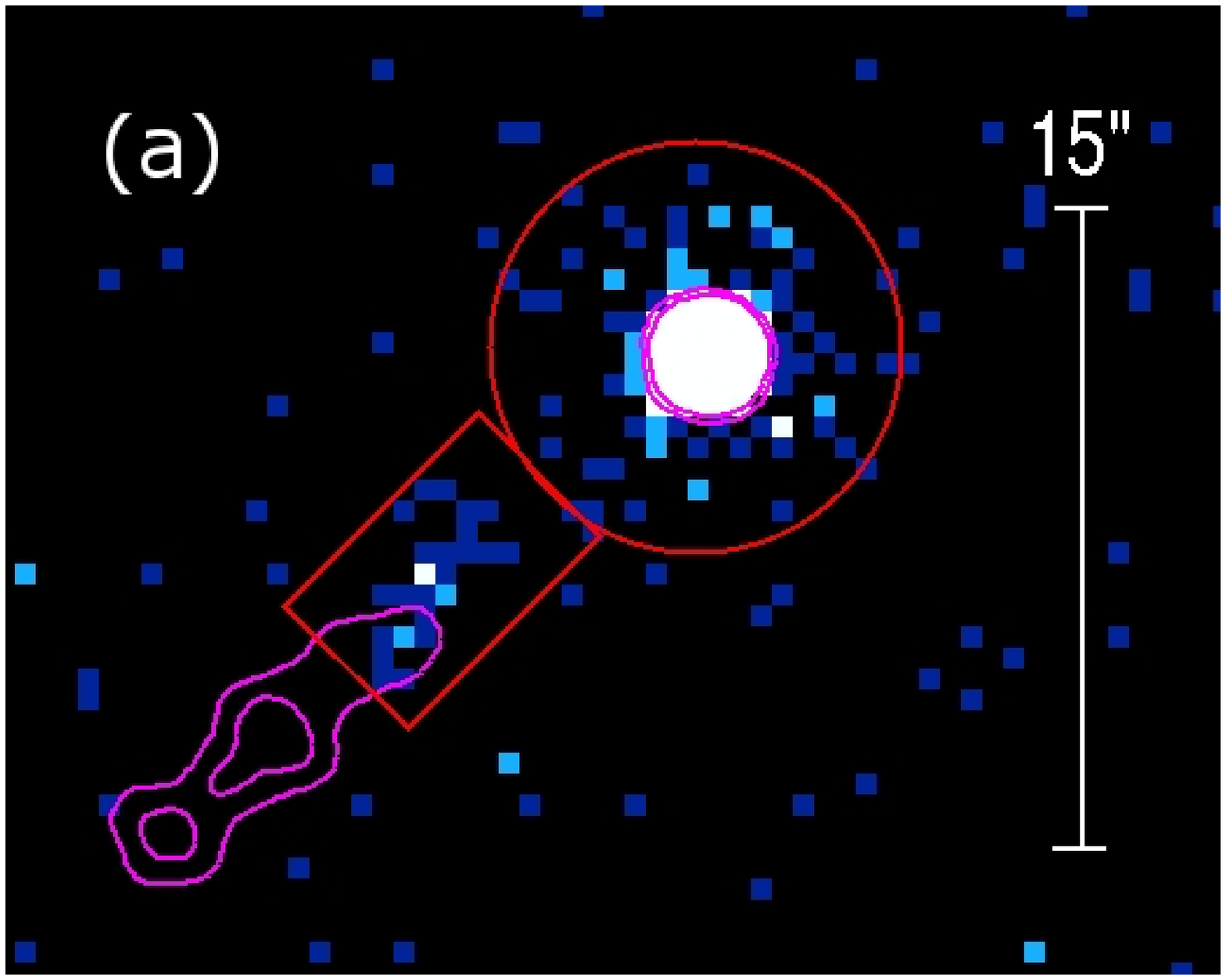}{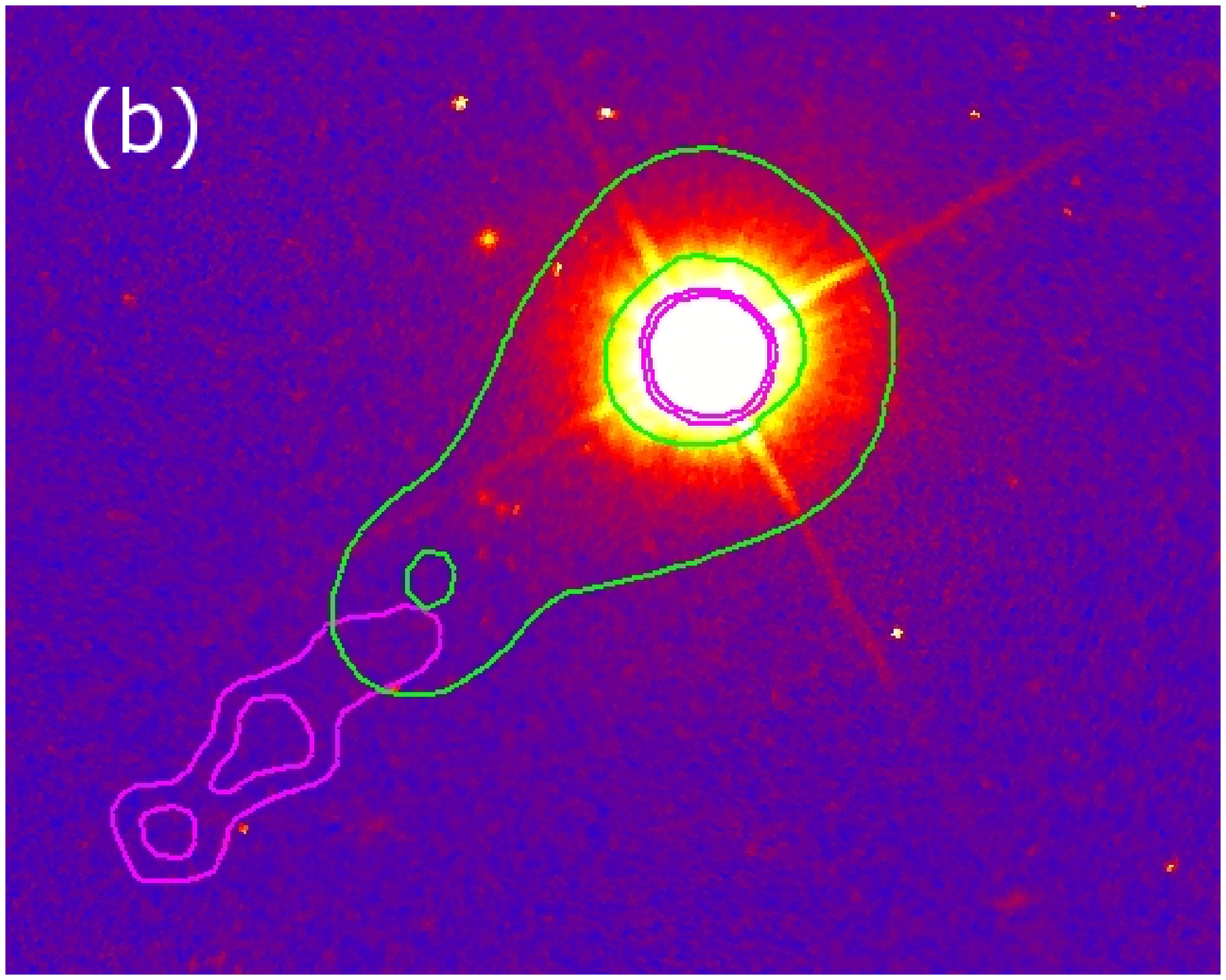}
\plottwo{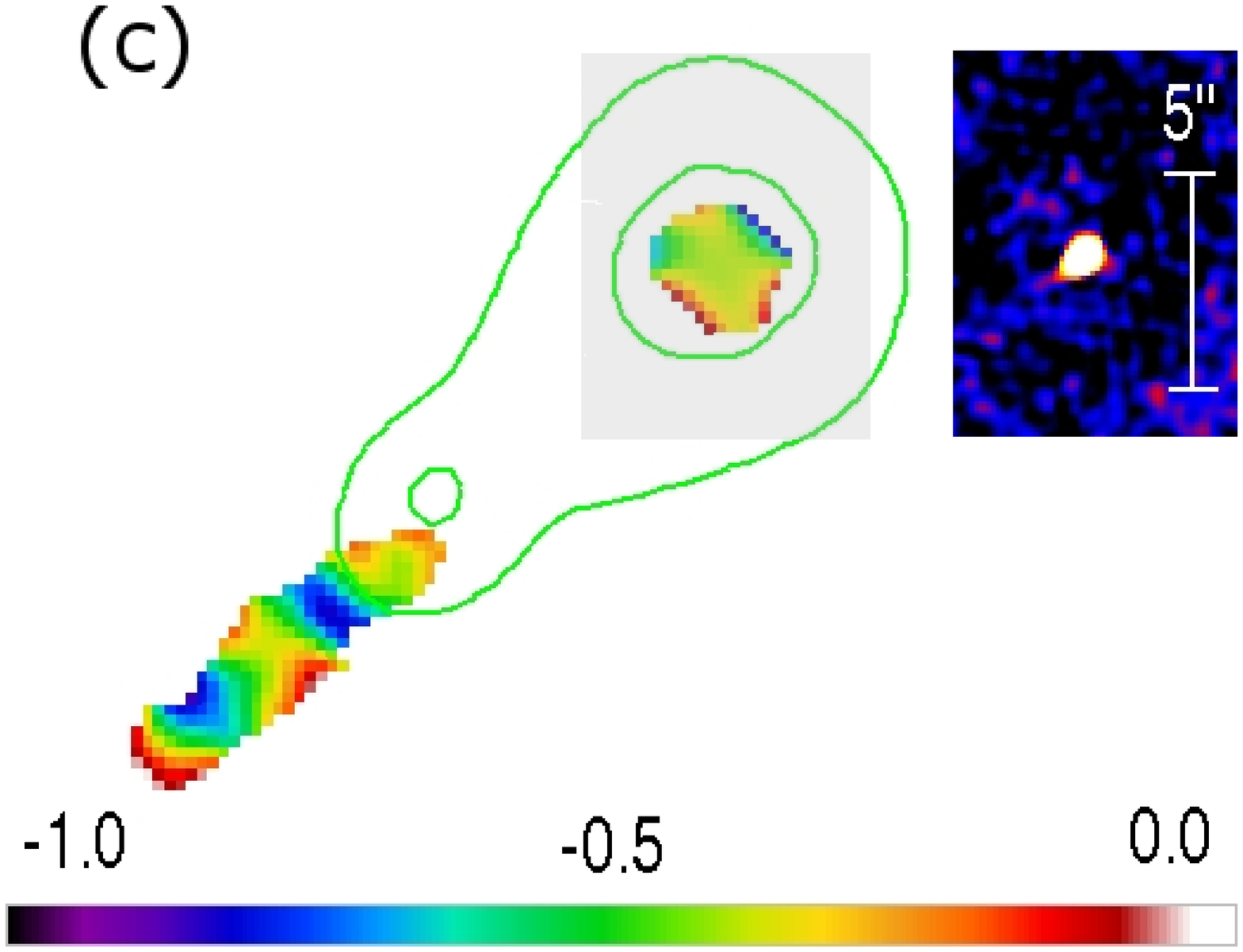}{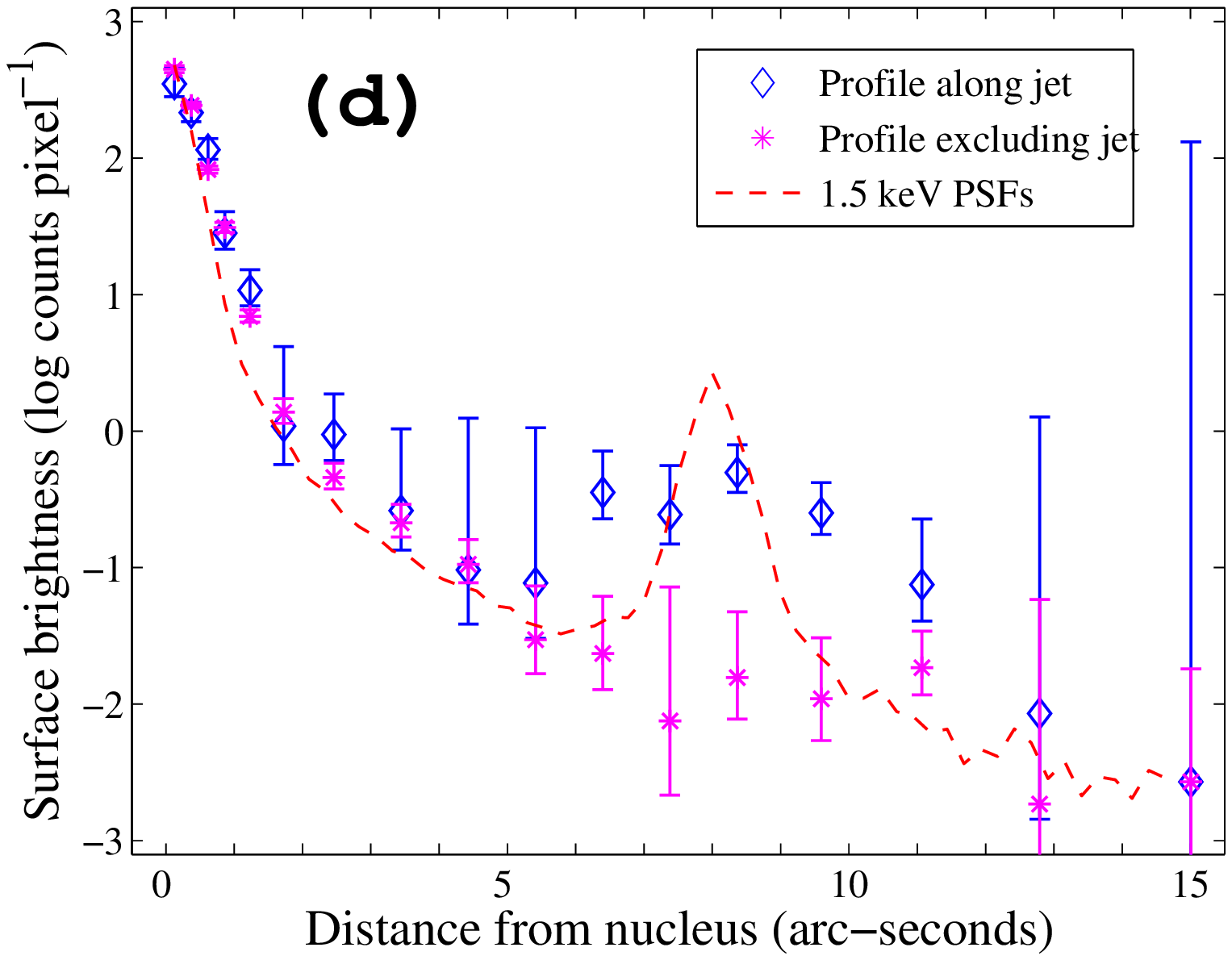}

\caption{(a) \textit{Chandra} 0.5--4~keV ACIS-S3 image (with pixel
randomization removed) of PG 1004+130. There are $\sim$30 counts in
the jet (box region) and $\sim$1600 in the nucleus (circular region)
in this band. Overlaid magenta contours show 4.9~GHz emission observed
by the {\it VLA} with 1.5$''$ resolution; the X-ray jet occurs
upstream of most of the radio emission. 15$''$ is 57 projected
kpc. (b) \textit{Hubble Space Telescope} WFPC2 image of PG~1004+130,
taken with the F606W filter, overlaid with green contours from the
smoothed \textit{Chandra} image and magenta radio contours duplicated
from (a). The scale is identical to (a) and the X-ray contour levels
are 0.05 and 0.20 counts~pixel$^{-1}$ (approximately 0.3 and 1.3
$\times$ 10$^{-33}$
erg~cm$^{-2}$~s$^{-1}~$Hz$^{-1}~$arcsec$^{-2}$). (c) Spectral-index
map of the FR~I jet (linear color scale given at bottom), generated
from 4.9 and 1.5 GHz {\it VLA} images of $\sim$1.5$''$
resolution. Inset shows a 0.5$''$ resolution 5 GHz image. (d)
Background-subtracted X-ray radial profiles of the nucleus and jet
compared with normalized 1.5~keV point spread functions located at the
nucleus and jet centroids. The nucleus is unresolved, but the jet is
extended. }
\end{figure}

\clearpage
\begin{figure}
\epsscale{0.7}
\plotone{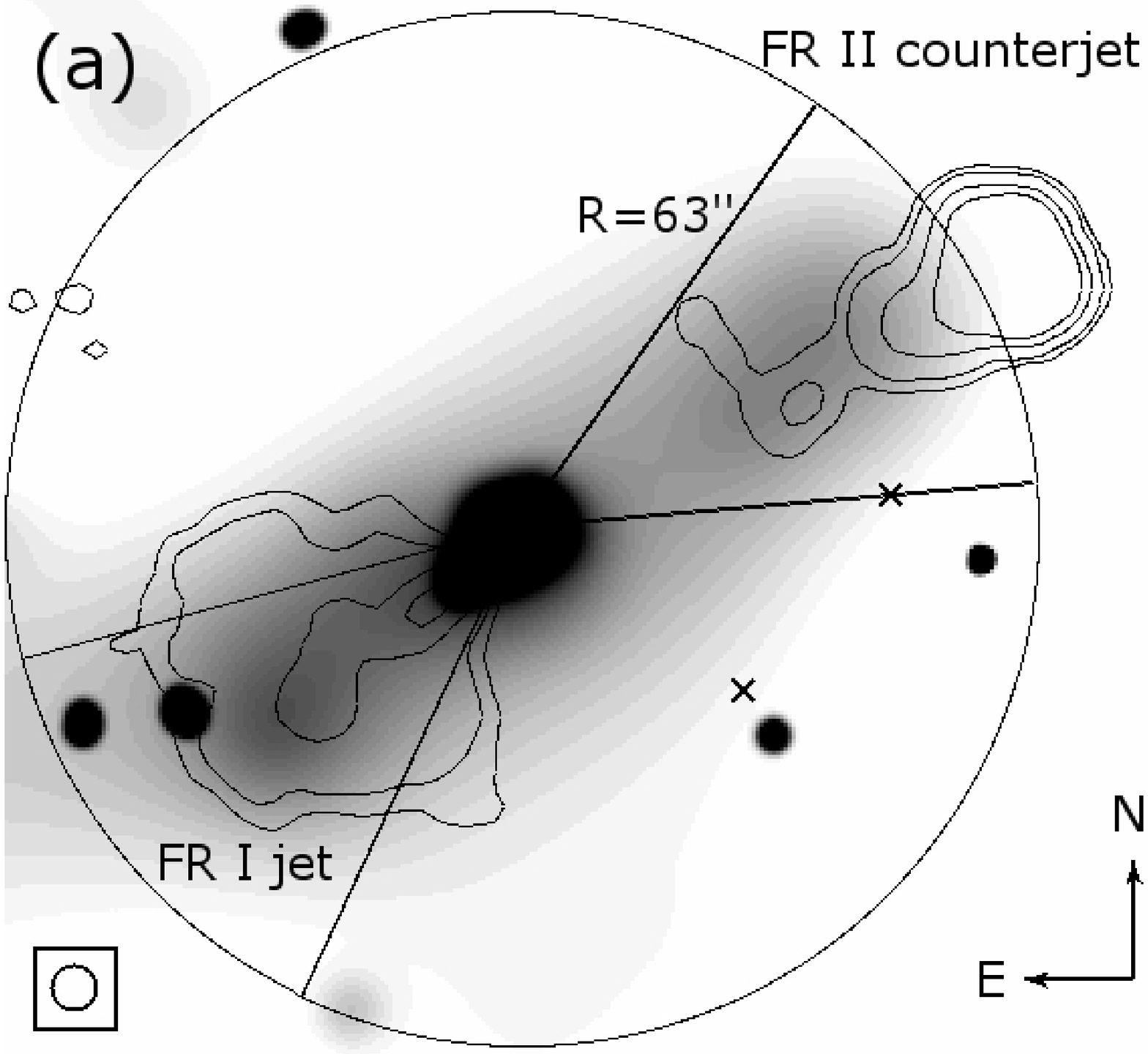}
\plotone{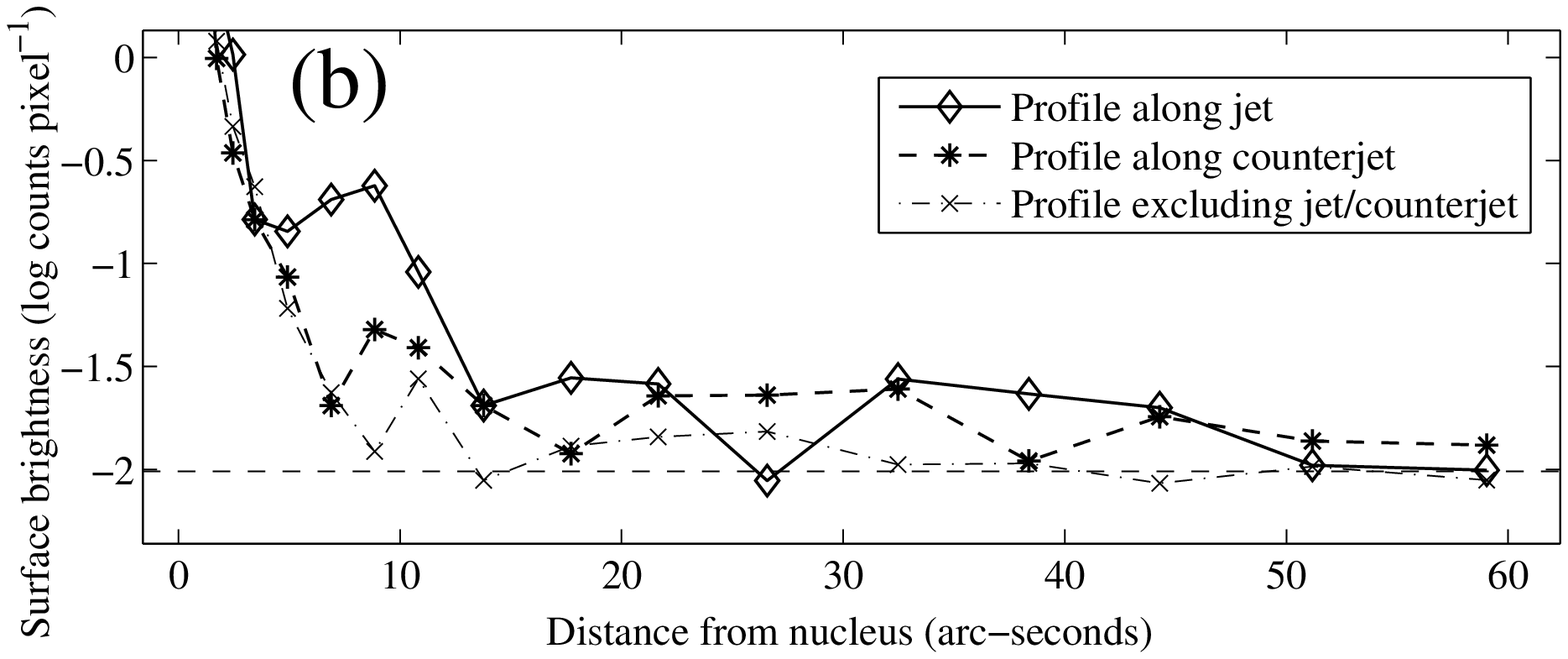}
\caption{(a) Adaptively smoothed 0.5--4~keV \textit{Chandra} image
overlaid with 1.4 GHz radio contours from the FIRST survey. Levels are
1.5, 3, 8, and 15 mJy beam$^{-1}$; the 5.4$''$ FWHM beam is shown at
lower left. There are two unrelated background X-ray sources located
near the end of the SE radio jet, and the positions of the two
galaxies believed to be associated with PG~1004+130 are indicated with
crosses. Diffuse X-ray emission is observed along both the SE jet and
the NW counterjet. (b) Surface-brightness radial profiles along the
jet, counterjet, and non-jet regions as indicated in (a);
contaminating point sources have been removed for this analysis, and
the local background is indicated with a dashed line. Past the inner
16$''$ where the X-ray jet is found, the surface brightness of the
diffuse emission is similar to the SE and the NW, and in both the jet
and counterjet regions the diffuse emission is significantly higher
than the background out to $\approx$50$''$.}
\end{figure}

\clearpage
\begin{figure}
\epsscale{1.1}
\plottwo{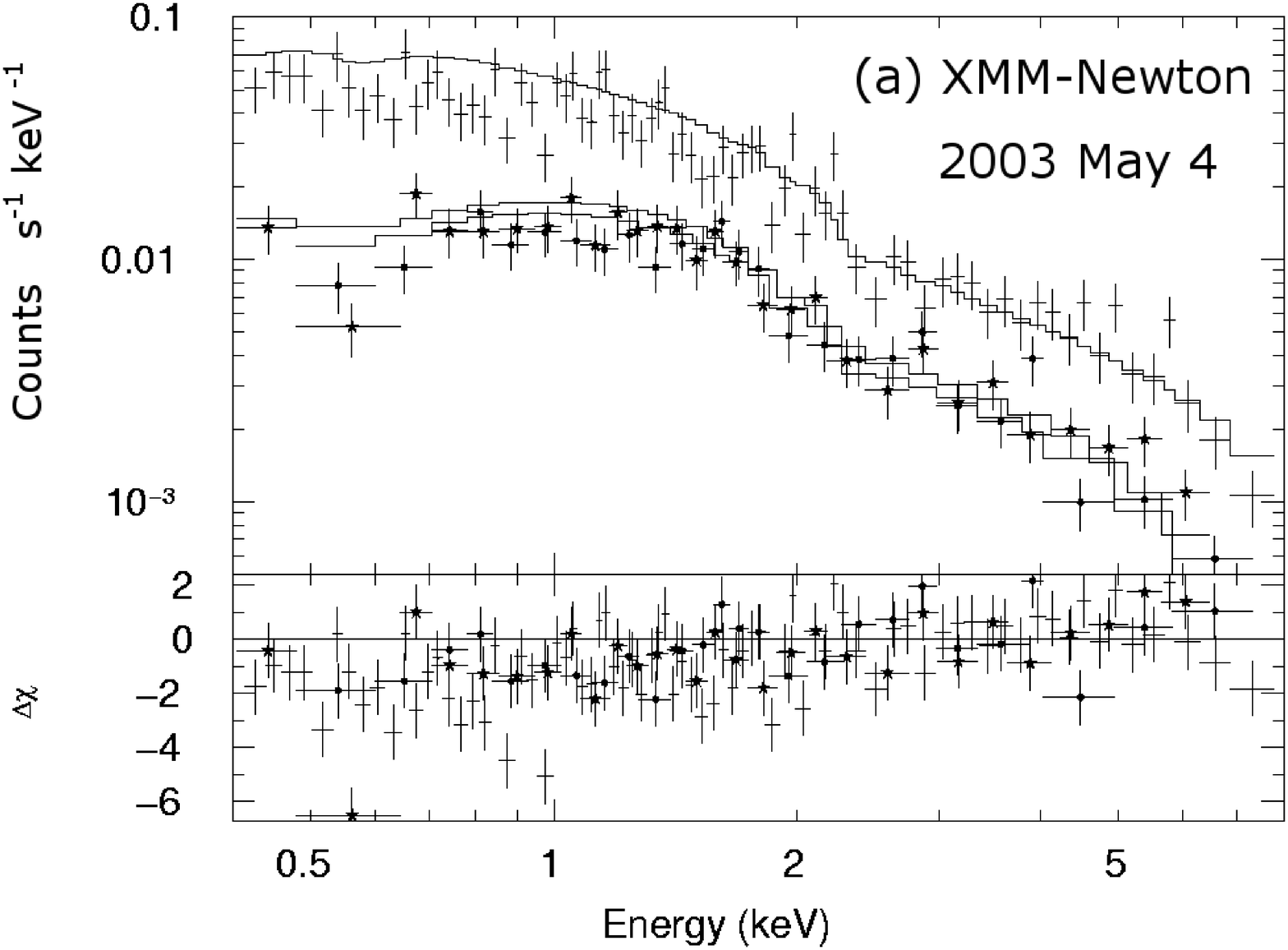}{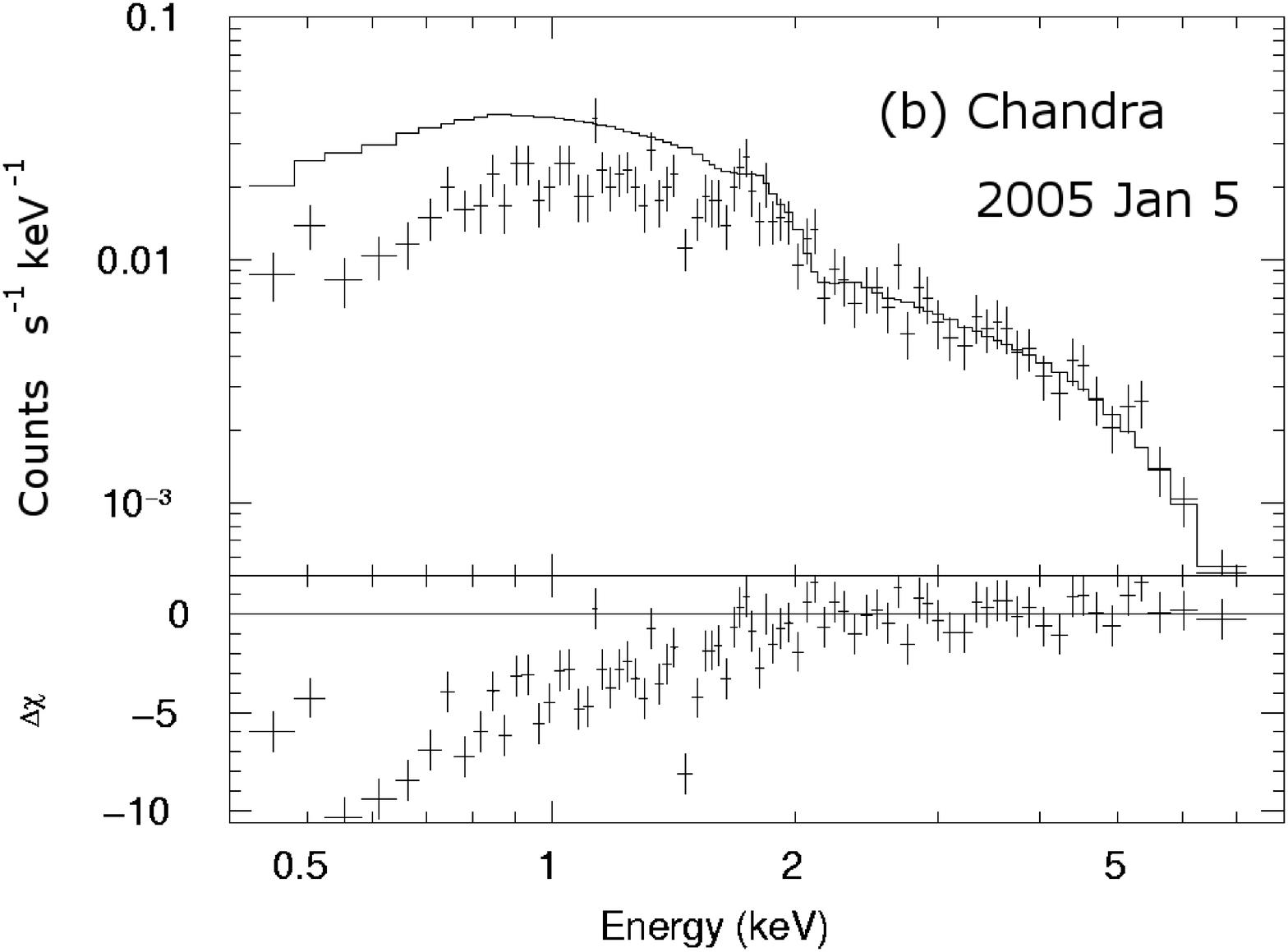}
\epsscale{1.1}
\plottwo{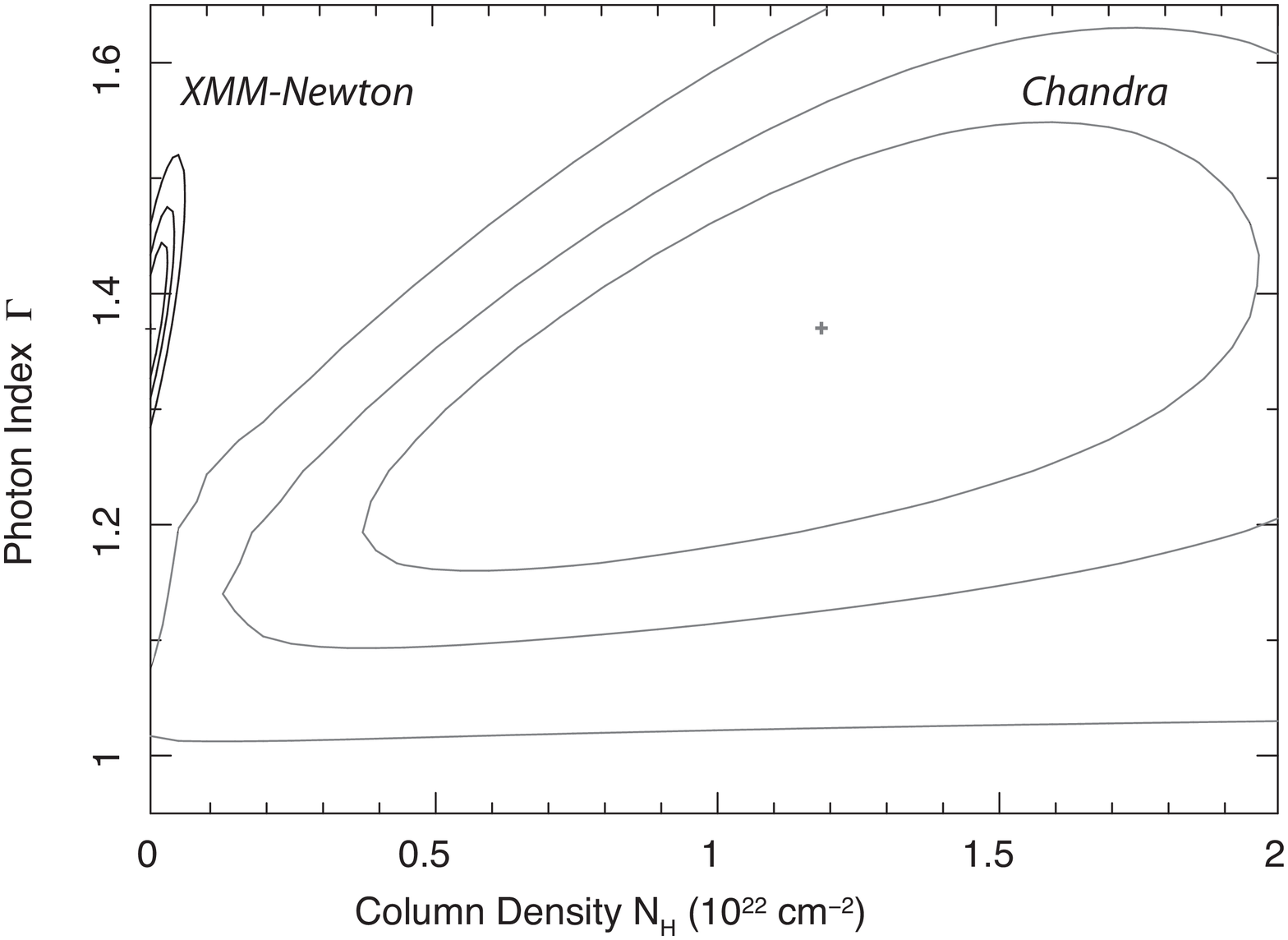}{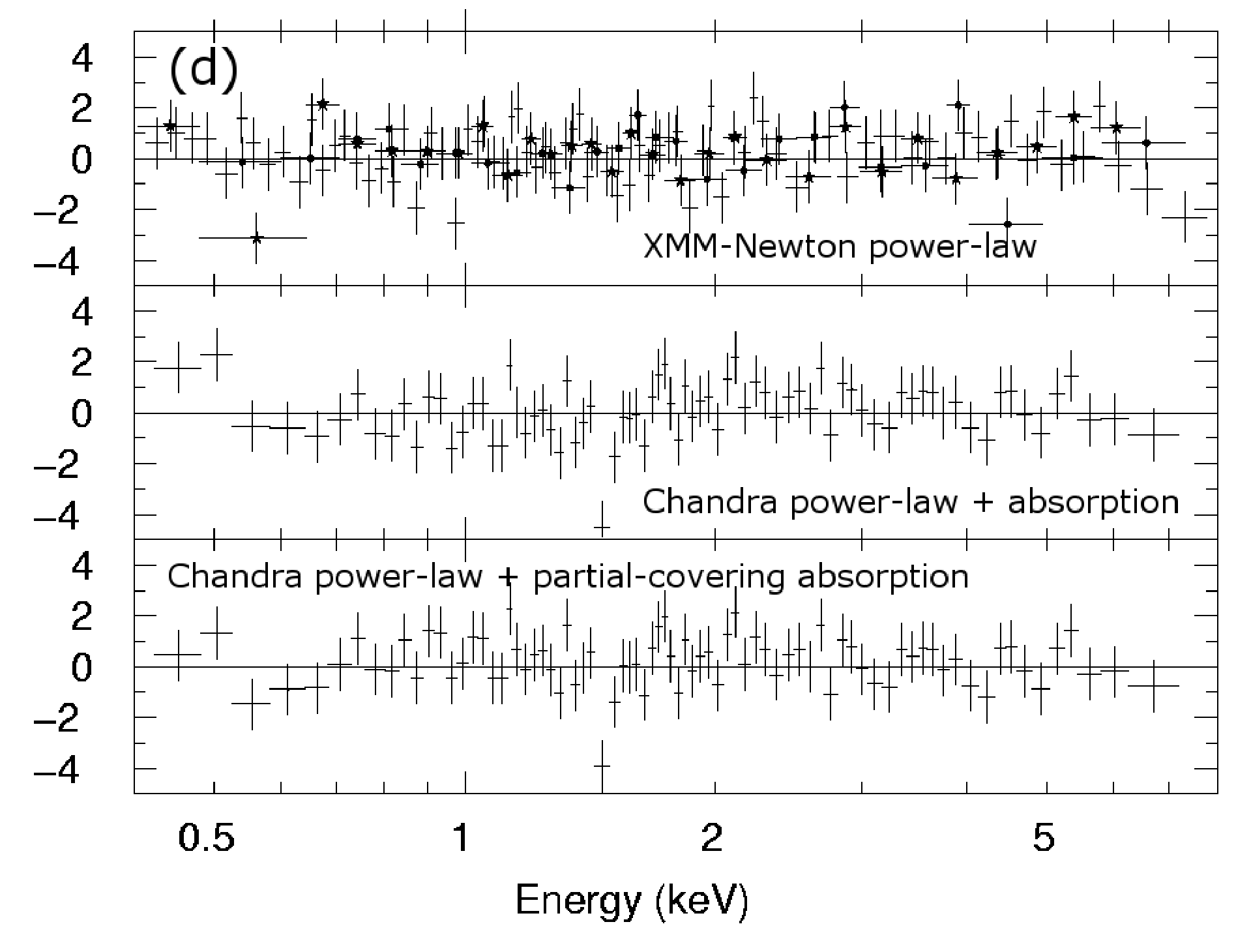}
\caption{(a) \textit{XMM-Newton} pn (plain crosses) and MOS (dots and
stars) spectra of PG~1004+130, shown with a model consisting of fixed
Galactic absorption and a power-law component; the power-law fit was
performed over the 2--8~keV range and then extrapolated to lower
energies. The residuals indicate the deviation of the data from the
model in units of $\sigma$, and reveal only minimal evidence for
intrinsic absorption. (b) \textit{Chandra} ACIS-S3 spectra analyzed as
above. It is apparent that the 2--8~keV fit does not satisfactorily
extend to the soft X-ray band, indicating significant absorption. The
photon index ($\Gamma \approx$ 1.5) is consistent with that of the
\textit{XMM-Newton} fit. (c) $N_{\rm H}-\Gamma$ contours (at 68\%,
90\%, and 99\% confidence) for the best-fit models illustrating that
the tight constraints on any intrinsic neutral absorption in the {\it
XMM-Newton} spectrum (black contours) conflict with the column density
of the partial-covering absorber in the {\it Chandra} spectrum (gray
contours). The photon indices are again similar but slightly flatter
($\Gamma \approx$ 1.4) than for 2--8 keV power-law models. (d)
Residuals from the fits, showing the {\it XMM-Newton} power-law model
(top), the {\it Chandra} power-law model with intrinsic neutral
absorption (middle), and the {\it Chandra} partial-covering absorption
model (bottom).}
\end{figure}

\clearpage
\begin{figure}
\epsscale{0.95}
\plotone{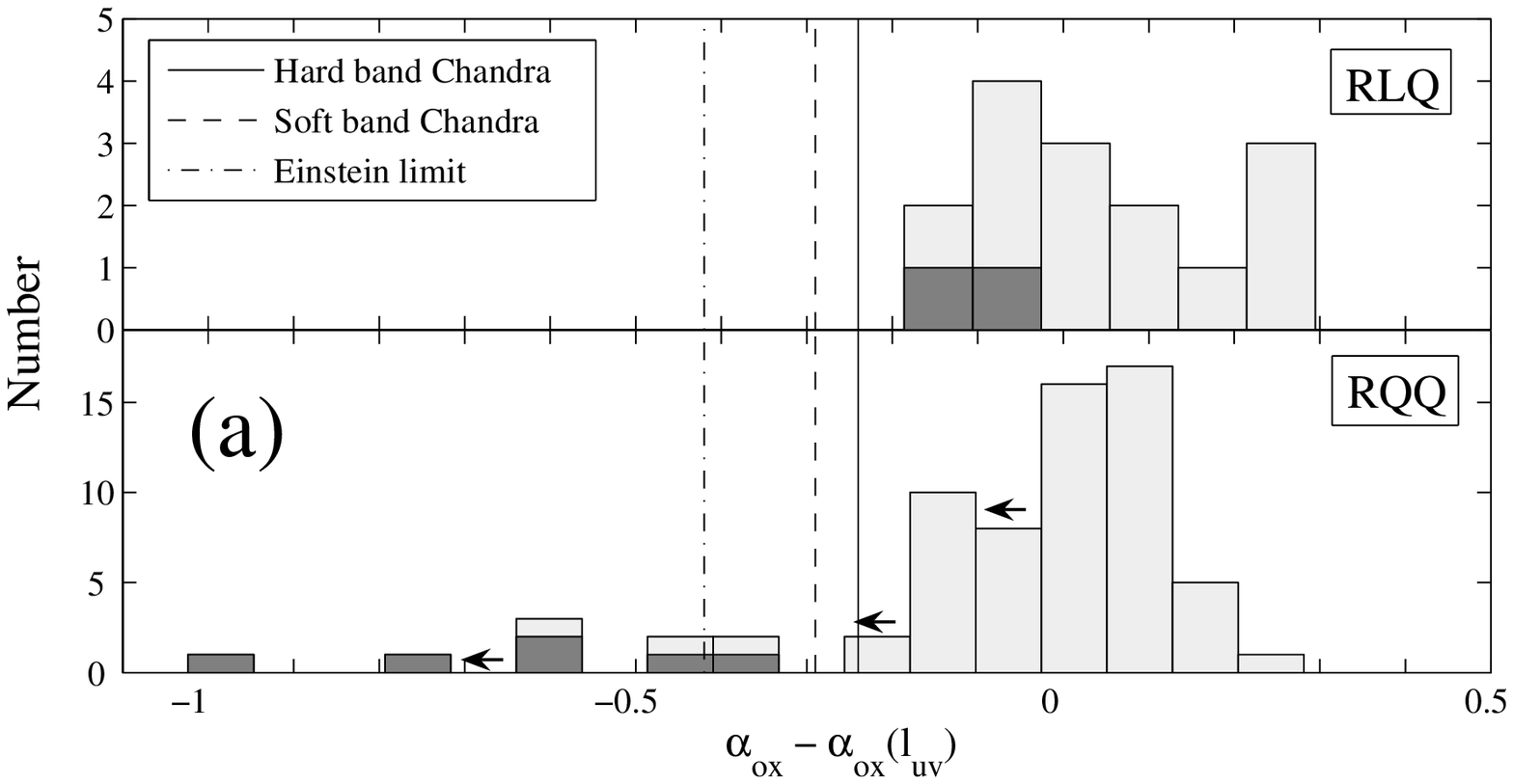}
\plotone{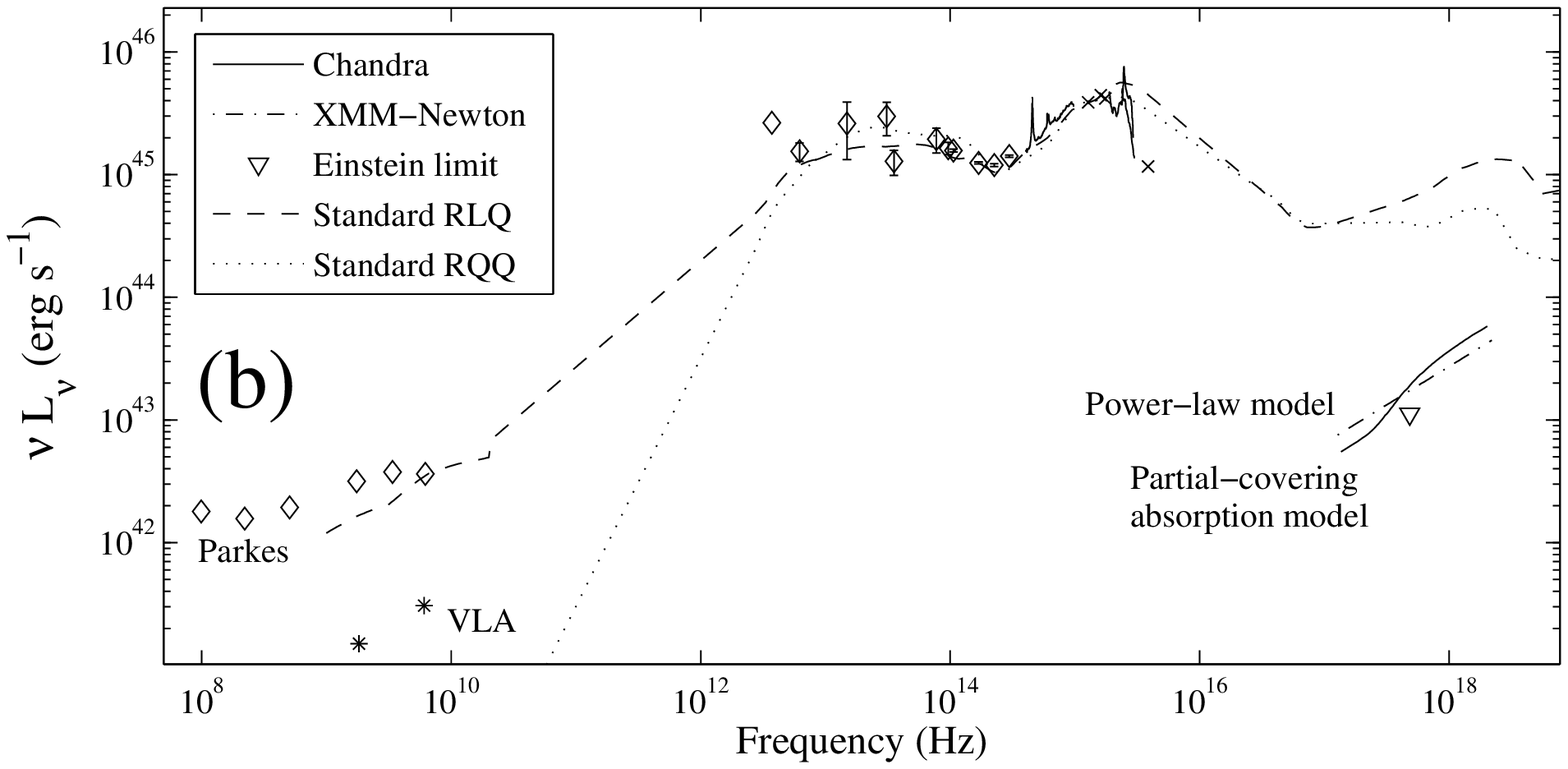}
\caption{(a) Comparison of the optical/UV-to-X-ray spectral slope,
${\alpha}_{\rm ox}$, of PG~1004+130 (corrected for luminosity
dependence following Steffen et al.~2006) with radio-loud (top) and
radio-quiet (bottom) quasars from the Palomar-Green survey. The subset
of quasars with intrinsic absorption is shaded dark and limits are
indicated with arrows. Three values of ${\alpha}_{\rm ox}$ are given
for PG~1004+130, one from the {\it Einstein} limit (dot-dashed line)
and two from the {\it Chandra} spectrum, the first based on the
partial-covering absorber model (``soft band'', dashed line) and the
next based on the 2--8~keV power-law model (``hard band'', solid
line). PG~1004+130 shows an anomalously steep decline in intensity
from optical/UV-to-X-ray wavelengths, an effect that is reduced for
the hard band. (b) Rest-frame SED for PG~1004+130, with standard RLQ
and RQQ SEDs from Elvis et al.~(1994) overplotted for reference. The
Parkes data include the extended radio emission, while the {\it VLA}
data are given for the nucleus alone. The best-fit ``unfolded'' models
for the {\it XMM-Newton} and {\it Chandra} spectra are shown. PG~1004+130
is X-ray weak relative to other RLQs. SED data and references are
listed in Table 1.}
\end{figure}

\begin{figure}
\epsscale{1.0}
\plotone{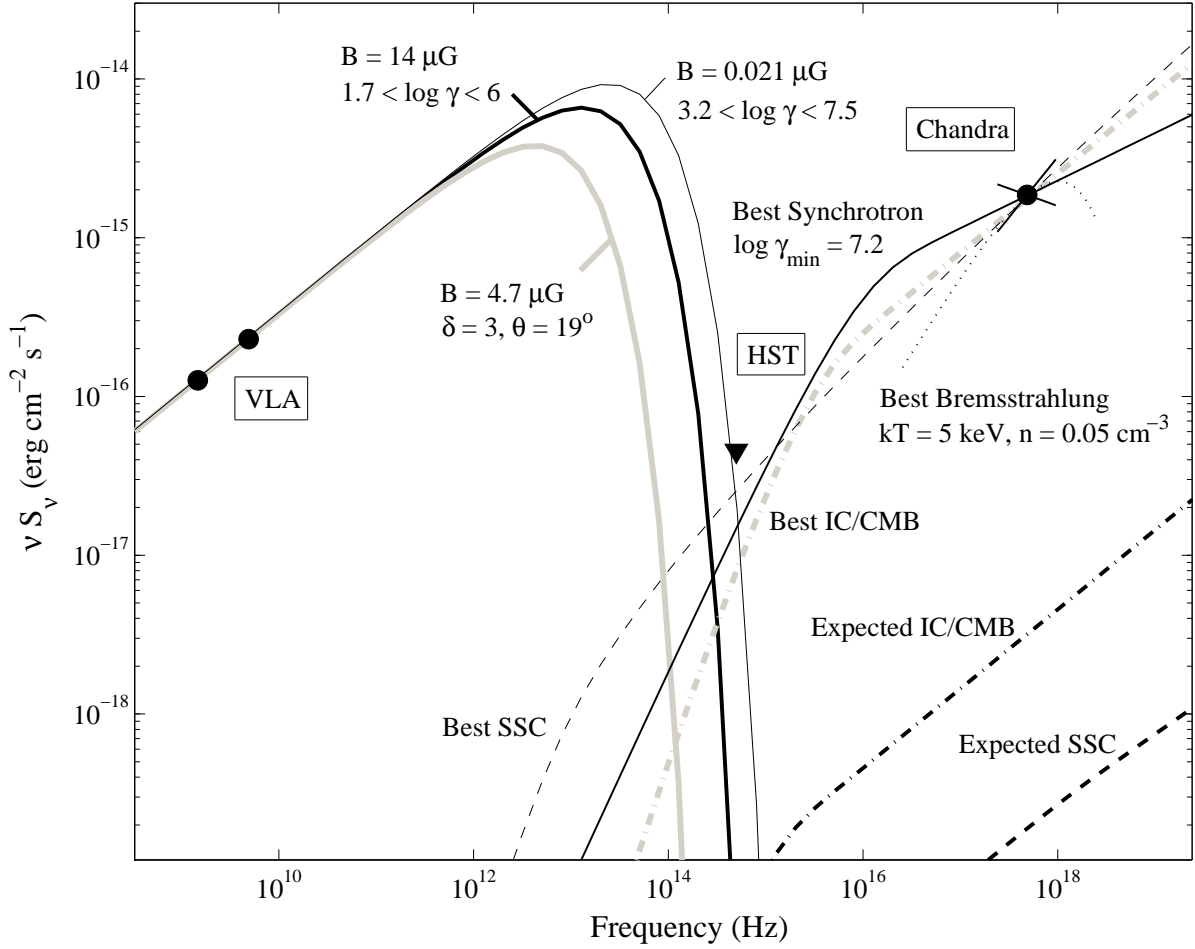}
\caption{Application of various models to the multiwavelength SED of
the PG~1004+130 jet. Solid lines show synchrotron components, dashed
lines show associated SSC emission, dot-dashed lines show associated
IC/CMB emission, and the dotted line shows a bremsstrahlung model. The
thickest black lines correspond to unbeamed models with an
equipartition magnetic field of $B_{eq}$=14 $\mu$G, the thin black
lines illustrate the ``best-case'' SSC model, with a highly
sub-equipartition magnetic field of $B$=0.021 $\mu$G, and the thick
gray lines illustrate the ``best-case'' IC/CMB model, in which the
line-of-sight angle is constrained to be less than 19$^{\circ}$ for
the required Doppler boosting of $\delta$=3. The X-ray photon index
predicted by the SSC and IC/CMB models is flatter than observed and
the presence of the X-ray emission largely upstream of the parent
synchrotron electrons is difficult to explain. We consider the
two-component synchrotron model to provide the most likely explanation
of the multiwavelength jet emission.}
\end{figure}

\end{document}